\documentclass[referee,aps,prd,floats,nofootinbib,longbibliography]{revtex4-1}
\usepackage{float}
\usepackage{tikz}
\usetikzlibrary{matrix}
\usepackage{graphicx}
\usepackage{graphics}
\graphicspath{{./figs/}}
\usepackage{bm}
\usepackage{amsmath,amssymb}
\usepackage{aas_macros}
\usepackage{calrsfs}
\usepackage{natbib}
\usepackage[colorlinks=true]{hyperref}
\usepackage{breqn}
\usepackage{color}
\usepackage{changes}
\usepackage[T1]{fontenc}
\usepackage{ae,aecompl}
\definechangesauthor[name={Leo}, color=orange]{LB}
\definechangesauthor[name={Agnes}, color=teal]{AF}
\definechangesauthor[name={Olivier}, color=magenta]{OM}
\definechangesauthor[name={Jacques}, color=red]{JL}
\definechangesauthor[name={Aurelien}, color=brown]{AH}

\usepackage{ulem}

\def\d{\mathrm{d}}
\def\e{\mathrm{e}}

\def\M{\mathcal{M}}
\def\({\left(}
\def\){\right)}
\def\[{\left[}
\def\]{\right]}

\def\aap{\textit{Astronomy and Astrophysics}}
\def\d {\mbox{d}}

\def\be{\begin{equation}}
\def\ee{\end{equation}}
\def\bea{\begin{eqnarray}}
\def\eea{\end{eqnarray}}

\usepackage{graphicx}	
\usepackage{amsmath}	
\usepackage{amssymb}	
\usepackage[draft]{fixme}
\fxsetup{layout=footnote, marginclue}

\begin{document}
\title{Constraining massless dilaton theory at Solar system scales with the planetary ephemeris INPOP}

\author{
L. Bernus $^{1,2}$,
O. Minazzoli $^{3}$,
A. Fienga $^{2,1}$,
A. Hees $^4$
M. Gastineau $^{1}$,
J. Laskar $^{1}$,
P. Deram $^2$,
A. Di Ruscio$^{2,5}$\\
$^{1}$IMCCE, Observatoire de Paris, PSL University, CNRS, Sorbonne Universit\'e, 77 avenue Denfert-Rochereau, 75014 Paris, France \\
$^{2}$G\'eoazur, Observatoire de la C\^ote d'Azur, Universit\'e C\^ote d'Azur, IRD, 250 Rue Albert Einstein, 06560 Valbonne, France\\
$^{3}$Artemis, Universit\'e C\^ote d'Azur, CNRS, Observatoire de la C\^ote d'Azur, BP4229, 06304, Nice Cedex 4, France\\
$^4$SYRTE,  Observatoire  de  Paris,  Universit\'e  PSL, CNRS,  Sorbonne  Universit\'e,  LNE,  61 avenue de l’Observatoire 75014  Paris,  France\\
$^{5}$ Dipartimento di Ingegneria Meccanica e Aerospaziale, Sapienza Universit\`a di Roma, via Eudossiana 18, 00184 Rome, Italy\\
}

\begin{abstract}
	We expose the phenomenology of the massless dilaton theory in the Solar system for a non universal quadratic coupling between the scalar field which represents the dilaton, and the matter. Modified post-Newtonian equations of motion of an $N$-body system and the light time travel are derived from the action of the theory. We use the physical properties of the main planets of the Solar system to reduce the number of parameters to be tested to 3 in the linear coupling case.
	In the linear case, we have an universal coupling constant $\alpha_0$ and two coupling constants $\alpha_T$ and $\alpha_G$ related respectively to the telluric bodies and to the gaseous bodies. 
	We then use the planetary ephemeris, INPOP19a, in order to constrain these constants. We succeeded to constrain the linear coupling scenario and the constraints read $\alpha_0=(1.01\pm23.7)\times 10^{-5}$, $\alpha_T=(0.00\pm24.5)\times 10^{-6}$, $\alpha_G=(-1.46\pm12.0)\times 10^{-5}$, at the 99.5 \% C.L. 
	\end{abstract}
\maketitle

\section{Introduction}
\label{sec:intro}

While the Equivalence Principle (EP) is at the heart of general relativity (GR), it has been argued that there actually exist no strong theoretical reasons to expect this principle to be valid in Nature, notably suggesting that GR should be replaced by a more accurate theory of relativity \cite{damour:2012cq}. This argument provides a strong motivation to search for an observational violation of the EP.

Amongst all the alternative theories of gravitation, scalar-tensor theories have been widely studied due to their simplicity as well as their manifest ability to give somewhat natural answers to apparently different issues in fundamental physics. In this class of theory, there is a priori no fundamental symmetry that can justify the EP to be valid. It is therefore natural to consider a non-minimal coupling between the scalar field and matter \cite{damour:2012cq}. Furthermore, scalar-fields with scalar-matter couplings are ubiquitous in several attempts to unify the whole fundamental interactions of physics \cite{damour:2012cq}, such as in superstring or Kaluza-Klein theories for instance. Such a non-minimal coupling is also considered in some models of Dark Matter and Dark Energy \cite{stadnik:2015yu,arvanitaki:2015qy}. In addition, it has been argued that scalar-tensor theories satisfying the EP at the classical level can exhibit a non-minimal coupling between the scalar field and matter due to quantum loop corrections \cite{armendariz2012prd}. This may actually lead to an impossible existence of a scalar-field minimally coupled to matter fields in Nature. Hence, all scalar-tensor theories are likely to violate the EP to some extent. In what follow, we shall generically name such a class of theories ``\textit{dilaton theory}'', in reference to the massless dilaton field that is generic in superstring theories, and which non-minimally couples to matter fields in the perturbative effective action \cite{damour1994npb,*Greene_2007,gerard:1995fk,*gerard:1997nr,*damour:2002vn,*kraiselburd:2018pr}.

From an experimental point of view, the violation of the EP has been mainly constrained by two types of experiments: (i) tests of the universality of free fall (UFF) and (ii) search for space-time variations in the constants of Nature. UFF is tested on Earth by measuring the relative acceleration between two test masses at the level of $10^{-13}$ using torsion balances \cite{schlamminger:2008zr,*wagner:2012fk}. Recently, it has also been tested in space with the MICROSCOPE experiment at the level of $10^{-14}$ \cite{touboul2019cqg}. Tests at the astronomical scale are performed in considering the free fall of the earth and Moon towards the Sun. These experiments give limits at the level of $10^{-13}$ as well \citep{williams2009}. These results have been interpreted in the context of dilaton theories \cite{berge:2018aa,*hees:2018aa,viswanathan2018mn}. Besides UFF, there exist typically three different types of searches for variations of the constants of Nature, all of them comparing the behavior of two collocated atomic clocks working on different atomic transitions \cite{uzan2011lrr}. First, comparing the long-term linear evolution of two atomic clocks gives a constraint on the cosmological evolution of the scalar field \cite{damour:2002vn,*damour:2002ys}. Such an experiment has been performed by several groups in the world \cite{guena:2012ys,*rosenband:2008fk,*leefer:2013xy,*godun:2014sf,*huntemann:2014nr} leading to constraints on a linear drift of several constants of Nature at the relative level of $10^{-16}$ per year, such as for the fine structure constant for instance. A second signature commonly searched for using atomic clocks comparison is a harmonic temporal evolution of the constants of Nature \cite{van-tilburg:2015fj,*hees:2016uq,*kennedy:2020aa,*savalle:2021wd} motivated mainly by models of ultralight Dark Matter \cite{stadnik:2015yu}. The last type of behaviors consists in searching for a relative variation of the constants of Nature as a function of the gravitational potential. Such a search has also been performed using atomic clocks \cite{peil:2013ul,*ashby:2018aa} but also using astrophysical observations \cite{hu:2020aa,*hees:2020aa}.

Given the ever increasing constraints on any EP violation from observations and experiments, this may seem to be a fatal blow for scalar-tensor theories in general, and to superstring theories in particular. Nevertheless,  several types of decouplings exist that can suppress EP violations below experimental limits---may they be dynamical \cite{damour1994npb,gerard:1995fk,*gerard:1997nr,*damour:2002vn,*kraiselburd:2018pr,khoury:2004fk,*khoury:2004uq,*Greene_2007,hinterbichler:2010fk,*hinterbichler:2011uq,vainshtein:1972ve,*babichev:2009fk,minazzoli:2013fk}, intrinsic \cite{ludwig:2015pl,minazzoli2016prd,minazzoli:2021cq} , or due to symmetries \cite{carroll:1998pl}.

Hence, it seems important to continue to explore the phenomenology of scalar-tensor theories with non-minimal couplings to matter and to confront it to observations.

While UFF has been tested to an exquisite level with the MICROSCOPE experiment, the latter actually constrain a very narrow region of the parameter space when interpreted in the framework of a dilaton theory \cite{berge:2018aa,*hees:2018aa}. This is due to the very specific composition of the free falling masses in the experiment. On the other hand, planets in the Solar system may help to further expand the region of the  parameter space being explored, due to their different compositions and scales. Nevertheless, while the dilaton theory has already been tested using atomic clocks, currently this theory has not been tested at the scale of the Solar system. In this manuscript, we will show that a non universally coupled massless dilaton induces a WEP violation that can be constrained in the Solar system. We will show that the physical nature of the Solar system allows to simplify the dilaton modelling and reduce the number of parameters to be constrained. The dilaton theory introduces only a limited number of fundamental parameters to be tested: 5 in the case of a linear coupling between the scalar field and matter and 10 in the case of a quadratic coupling. Nevertheless, we will show that in the Solar system, these fundamental parameters appear as combinations such that the number of coefficients to be tested can be reduced to 3 in the linear coupling scenario. These 3 parameters are derived constant from the fundamental coupling constants of the theory. This work is a first step to test dilaton theory in the Solar system.  Finally, we will present a data analysis of the recent planetary ephemeris INPOP19a that leads to a constraint on the dilaton parameters in the case of a non universal linear coupling between the dilaton field and the matter fields.

In Sec. \ref{sec_pheno}, we summarize the phenomenology induced by a massless dilaton within the Solar system. We present the expression of the action, of the equations of motion for a $N$-body system and also the expression of the light time travel. The details of the mathematical derivations are provided in Appendix \ref{ap_pncalcul}, and some complements about the Lagrangian and Hamiltonian frameworks and about the first integrals of the equations of motion in the massless dilaton theory are given in Appendix \ref{ap_lagr}. In Sec. \ref{sec_num}, we expose the numerical methods used in our analysis. First we show how the number of parameters of the dilaton theory can be efficiently reduced and present how we implemented the modified equations of motion to our Solar system planetary ephemeris INPOP19a to build a test of the  theory of massless dilaton. We also expose the statistical criterion used to constrain the parameters of the theory. In section \ref{sec_resi}, we present the residuals obtained as a function of the dilaton parameters and deduce a likelihood distribution of these parameters. In Sec. \ref{sec_resu}, we deduce from the likelihood distributions our results for a linear coupling between the dilaton and the matter. These results are the posterior distributions of the parameters to be tested approximated by histograms. We also propose confidence intervals at 90\% Confidence Limit (C.L.) and at 99.5\% C.L. 
	Finally, we summarize our results in the conclusion (Sec. \ref{sec_concl}).

\section{Phenomenology of a massless dilaton in the Solar system}
\label{sec_pheno}
	\subsection{Action of the theory and field equations}
		The difference between Einstein's theory of gravitation and massless dilaton theory consists in the existence of a light scalar field $\varphi$ coupled to gravity field and matter  \cite{damour2010prd,minazzoli2016prd}. In the following, the term ``dilaton'' is identified with this scalar field. The massless dilaton theory contains four terms in its action. The first term consists in the Einstein-Hilbert action coupled to a differentiable function of the scalar field $f(\varphi)$, a kinetic term for the scalar field, the standard model Lagrangian for the matter fields, and a Lagrangian which parametrizes the interaction between the dilaton and matter. Let us consider a 4 dimensional manifold $\M$ described by a map denoted generically $(x^\mu)$ (we identify the coordinates chart and the map of $\M$). This action reads \cite{minazzoli2016prd}
		\begin{align}
		 	S[\bm{g},\psi_i,\varphi]&=\frac{1}{2\kappa c}\int \(f(\varphi) R -\frac{\omega(\varphi)}{\varphi}\varphi^{,\mu}\varphi_{,\mu}\)\sqrt{-g}\;\d^4x\nonumber\\
		 	&\quad+\frac{1}{c}\int\(\mathcal{L}_{SM}[\bm{g},\psi_i]+\mathcal{L}_{int}[\bm{g},\psi_i,\varphi]\)\sqrt{-g}\;\d^4x, \label{eq:action}
		\end{align} 
		where $\bm{g}$ is the metric tensor, $g$ is the determinant built with the matrix of the covariant components $g_{\mu\nu}$ of the metric tensor $\bm{g}$ in the map $(x^\mu)$, $\varphi$ is the scalar field named ``dilaton'', $f$ and $\omega$ are two twice differentiable functions of one variable, $\mathcal{L}_{SM}$ is the Lagrangian density of  matter described by the standard model, $\mathcal{L}_{int}$ is the Lagrangian density of the interaction between the dilaton and matter, $\psi_i$ represents the different matter fields. We could perform all the calculations with this action but the field equations and the derivation of the post-Newtonian equations of motion can be simplified by performing a conformal transformation. Let us introduce a conformal transformation of the metric tensor, $\bm{g}\mapsto\bm{g}^*(\bm{g},\varphi)$ 
		\begin{equation}
			g_{\mu\nu}=\frac{f_0}{f(\varphi)}g^*_{\mu\nu} , 
		\end{equation}
		and a transformation of the scalar field $\varphi\mapsto\phi(\varphi)$ which satisfies
		\begin{equation}
			\(\frac{\d \phi}{\d\varphi}\)^2=\frac{Z(\varphi)}{2}=\frac{\omega(\varphi)}{\varphi f(\varphi)}+\frac{3}{2}\(\frac{f'(\varphi)}{f(\varphi)}\)^2, \label{eq_changement_scalaire}
		\end{equation}
		where $f_0$ denotes the value of $f(\varphi)$ when $\varphi$ takes its background value $\varphi_0$ at infinite distance (a possible cosmological evolution of this background value is neglected in the present work). The map $(x^\mu)$ remains the same. Then a straightforward calculation shows that in these new variables, the action reads \cite{damour1992cqg,minazzoli2016prd}
		\begin{align}
			\tilde{S}[\bm{g}_*,\psi_i,\phi]&=S[\bm{g}(\bm{g}_*(\varphi(\phi))),\psi_i,\varphi(\phi)]\nonumber\\
			&=\frac{1}{2\kappa_*c}\int\(R_*-2g_*^{\mu\nu}\phi_{,\mu}\phi_{,\nu}\)\sqrt{-g_*}\ \d^4x\nonumber\\
			&\quad + \frac{1}{c}\int \mathcal{L}^*_m[\bm{g}_*,\psi_i,\phi]\sqrt{-g_*}\,\d^4x,
		\end{align}
		where
		\begin{equation}
		    \mathcal{L}^*_m[\bm{g}_*,\psi_i,\phi]=(\mathcal{L}_{SM}[\bm{g}(\bm{g}_*),\psi_i]+\mathcal{L}_{int}[\bm{g}(\bm{g}_*),\psi_i,\varphi(\phi)])\frac{\sqrt{-g}}{\sqrt{-g^*}},
		\end{equation}
		$R_*$ is the Ricci scalar computed with the new metric tensor $\bm{g}_*$, $g_*$ is the determinant computed with the covariant components $g^*_{\mu\nu}$ of the metric tensor $\bm{g}_*$ and 
		\begin{equation}
		\kappa_*=\frac{\kappa}{f_0}\, .
		\end{equation}
		The frame used after this transformation is usually called ``Einstein frame''. In this frame, a straightforward calculation shows that the field equations read \cite{damour1992cqg,minazzoli2016prd}
		\begin{equation}
			R^*_{\mu\nu}-\frac{1}{2}R_*g^*_{\mu\nu}=\kappa_*T^*_{\mu\nu}+2\phi_{,\mu}\phi_{,\nu}-g^*_{\mu \nu} \phi^{,\alpha}\phi_{,\alpha}\label{eqpremierjeueqf}
		\end{equation}
		\begin{equation}
			\Box_*\phi=-\frac{\kappa}{2}\frac{\partial(\mathcal{L}^*_{m})}{\partial \phi} \label{eqchampscaldil}
		\end{equation}
		where 
		\begin{equation}
			T^*_{\mu\nu}=-\frac{2}{\sqrt{-g_*}}\frac{\partial (\mathcal{L}_m\sqrt{-g_*})}{\partial g_*^{\mu\nu}}
		\end{equation}
		and $\Box_*=\nabla^*_\mu\nabla_*^\mu$ where $\bm{g}_*$ is used to compute the covariant derivatives.

	\subsection{Interaction between matter and the dilaton field}
	\label{subsec_intmatdil}
		An effective Lagrangian describing the interactions between matter and the dilaton is assumed to be described by some differentiable functions of the dilaton multiplied by the main matter fields \cite{minazzoli2016prd}
		\begin{align}
			\mathcal{L}_{int}&=\frac{D_e(\varphi)}{4e^2}F_{\mu\nu}F^{\mu\nu}-\frac{D_g(\varphi)\beta_3(g_3)}{2g_3}G^a_{\mu\nu}G_a^{\mu\nu}\nonumber\\
			&\quad-\sum_{i=e,u,d}\(D_{m_i}(\varphi)+\gamma_{m_i}D_g(\varphi)\)m_i\bar{\psi}_i\psi_i \label{eq_lagrdilaton}
		\end{align}
		where $F_{\mu\nu}$ is the Faraday tensor, $G^a_{\mu\nu}$ is the gluons tensor, $g_3$ is the strong force coupling constant, $\beta_3(g_3)=\mu\partial\ln g_3/\partial\mu$ is its beta function relative to the quantum scale invariance violation, $\mu$ is the energy scale of the relevant physical processes, $m_i$ is the fermions mass, $\psi_i$ their spinor, and $\gamma_{m_i}=-\mu\partial\ln m/\partial \mu$ is the beta function relative to the dimensional anomaly of the fermions masses coupled to the gluons.
		The $D_i(\varphi)$ functions describe the different couplings between the matter fields and the dilaton. $D_e$ characterizes the $\varphi$ dependency of the fine structure constant, $D_g$ characterizes the $\varphi$ dependency of the QCD mass scales $\Lambda_3$ and $D_{m_i}$ characterizes the quarks masses. The Lagrangian density \eqref{eq_lagrdilaton} is a straightforward non-linear generalisation of Damour \& Donoghue theory \cite{damour2010prd}. The theory considered here becomes equivalent to the one of Damour \& Donoghue \cite{damour2010prd} if we set $D_i(\varphi)=d_i\varphi$. Indeed, the dependency of the constant of Nature to the scalar field reads \cite{damour2010prd}
		\begin{subequations}
		\begin{align}
			\alpha(\varphi)&=(1+D_e(\varphi))\alpha \, , \\
			\Lambda_3(\varphi)&=(1+D_g(\varphi))\Lambda_3 \, , \\
			m_e(\varphi)&=(1+D_{m_e}(\varphi))m_e \, , \\
			m_q(\varphi)&=(1+D_q(\varphi))m_q,\quad q=u,d \, .
		\end{align}
		\end{subequations}
		Damour \& Donoghue \cite{damour2010prd} have shown that the matter action of a point mass, at a macroscopic level, becomes
		\begin{equation}
			S_m=-c^2\int m_A(\varphi)\d\tau_A
		\end{equation}
		where $A$ is the label of the considered body, $\d\tau_A=\sqrt{-\bm{g}\(\overrightarrow{\bm{v}_A},\overrightarrow{\bm{v}_A}\)}\d t$ where $\overrightarrow{\bm{v}_A}=c\overrightarrow{\bm{\d} z_A}/\d x^0$ the 4-velocity of body $A$ ($z_A\in\M$ is the position of body $A$ in the manifold and its coordinates in a generic map $(x^\mu)$ are $z^\mu_A$. The velocity vector in the tangent space $T_{z_A}\M$ is denoted $\overrightarrow{\bm{v}_A}$, and its coordinates are $v_A^\alpha=\d z_A^\alpha/\sqrt{-g_{\mu\nu} \d z_A^\mu \d z_A^\nu}$. The $\varphi$ dependency of $m_A$ depends on the internal structure of body $A$ and is responsible for the violation of the WEP. All this violation can be encoded in a coupling parameter
		\begin{equation}
			\alpha_A(\varphi)=\frac{\d\ln m_A}{\d \varphi}.
		\end{equation}
		Damour \& Donoghue have found a semi-analytical expression of $\alpha_A$ with respect to its atomic composition \cite{damour2010prd}. The non-linear generalisation is straightforward. At first order, one needs to replace the $d_i$ of Damour \& Donoghue by $D_i'(\varphi)$. It is convenient to decompose $\alpha_A$ into a universal term $\alpha_u$ (which does not depend explicitly on the atomic composition of the various bodies) and a non-universal part $\bar \alpha_A$ (which depends explicitly on the atomic composition). The coupling function reads
		\begin{equation}
			\alpha_A(\varphi)=\alpha_u(\varphi)+\bar{\alpha}_A(\varphi)\label{eq_couplage_premier}
		\end{equation}
		where a straightforward non linear generalisation of Eq. (71)-(72) of \cite{damour2010prd} reads 
		\begin{align}
				\alpha_u(\varphi)&=D_g'(\varphi)+[9.3\times 10^{-2}(D_{\hat{m}}'(\varphi)-D_g'(\varphi))\nonumber\\
				&\quad-1.4\times 10^{-4}(D_{m_e}'(\varphi)-D_g'(\varphi)]\label{eq_alphaudef}
		\end{align}
		and
		\begin{align}
				\bar{\alpha}_A(\varphi)&=(D_{\hat{m}}'(\varphi)-D_g'(\varphi))Q_{\hat{m}}^A+(D_{\delta m}'(\varphi)-D_g'(\varphi))Q_{\delta m}^A\nonumber\\
				&\quad+(D_{m_e}'(\varphi)-D_g'(\varphi))Q_{m_e}^A+D_e'(\varphi)Q_e^A \label{eq_couplnonuniv}
		\end{align}

		The coupling functions to the quarks have been redefined
		\begin{equation}
			D_{\hat{m}}(\varphi)=\frac{m_uD_{m_u}(\varphi)+m_dD_{m_d}(\varphi)}{m_u+m_d}
		\end{equation}
		\begin{equation}
			D_{\delta m}=\frac{m_dD_{m_d}(\varphi)-m_uD_{m_u}(\varphi)}{m_d-m_u}
		\end{equation}
		In Eq. \eqref{eq_couplnonuniv}, the \emph{dilatonic charges} appear: $Q_{\hat{m}}^A$, $Q_{\delta m}^A$, $Q_{m_e}^A$, and $Q_e^A$. They are responsible of the weak equivalence principle violation, because they depend explicitly on the atomic composition of body $A$. Charges $Q_{\hat{m}}^A$ and $Q_{\delta m}^A$ quantify the coupling between the quarks of body $A$ and the dilaton field, $Q_{m_e}$ the coupling with the mass of the electrons, and $Q_e$ the coupling with the electromagnetic field. Let $\mathcal{A}$ be the average number of nucleons of body $A$, and $\mathcal{Z}$ its average number of protons. Then the dilatonic charges are the same as the one in Damour \& Donoghue theory \cite{damour2010prd}: 
		\begin{equation}
			Q_{\hat{m}}^A=\frac{-3.6\times 10^{-2}}{\mathcal{A}^{1/3}}-2.0\times 10^{-2}\frac{(\mathcal{A}-2\mathcal{Z})^2}{\mathcal{A}^2}-1.4\times 10^{-4}\frac{\mathcal{\mathcal{Z}}(\mathcal{\mathcal{Z}}-1)}{\mathcal{A}^{4/3}} \label{eq_qhm}
		\end{equation}

		\begin{equation}
			Q_{\delta m}^A=1.7\times 10^{-3}\frac{\mathcal{A}-2\mathcal{Z}}{\mathcal{A}} \label{eq_qdm}
		\end{equation}
		\begin{equation}
			Q_{m_e}^A=5.5\times 10^{-4}\frac{\mathcal{Z}}{\mathcal{A}} \label{eq_qme}
		\end{equation}
		\begin{equation}
			Q_e=8.2\times 10^{-4}\frac{\mathcal{Z}}{\mathcal{A}}+7.7\times 10^{-4}\frac{\mathcal{Z}(\mathcal{Z}-1)}{\mathcal{A}^{4/3}} \label{eq_qe}
		\end{equation}
		
		Note that compared to Damour \& Donoghue's work \cite{damour2010prd}, we have removed the constant part of the dilatonic charges, because they are taken into account in $\alpha_u$, the universal coupling constant (see Eq. \eqref{eq_alphaudef}).

		We have computed some dilatonic charges by estimating the composition of the main bodies of the solar system. We report the values in Table \ref{tabchargdilatom}. Damour \& Donoghue have already computed these charges \cite{damour2010prd,damour2011cqg}.

		Let us note that if we follow the work of Nitti \& Piazza \cite{nitti2012prd}, the electromagnetic interaction should present a trace anomaly similarly to the other interactions and we should replace $D_e'(\varphi)$ by $D_e'(\varphi)-D_g'(\varphi)$ in Eq.  \eqref{eq_couplnonuniv}. In this case, if the coupling is universal, which means if it appears as $\mathcal{L}_{int}=D(\varphi)T_{SM}$---where $T_{SM}$ is the whole trace anomaly of the Standard Model---, then all the coupling functions are equal and we get $\bar{\alpha}_A=0$ such that the weak equivalence principle is still satisfied. Damour \& Donoghue don't take the electromagnetic trace anomaly into account, therefore in their theory, even with a universal coupling, the weak equivalence principle is broken. In our computations, it is always possible to replace $D_e'(\varphi)$ by $D_e'(\varphi)-D_g'(\varphi)$ if needed. In terms of Solar system phenomenology, if the coupling functions are different, then it changes nothing for testing alternative theories in an ``agnostic'' way, which are blind with respect to the choices of the definitions of coupling functions. Indeed, since we do not know anything about any coupling functions \emph{a priori}, constraining $D_e'(\varphi)$ instead of $D_e'(\varphi)-D_g'(\varphi)$ is exactly the same when we perform experimental tests.

		In our non-linear dilaton theory, some second order terms will appear at the post-Newtonian order. We introduce the quadratic coupling function
		\begin{equation}
			\beta_A(\varphi)=\frac{\d\alpha_A}{\d\varphi}=\frac{\d^2\ln m_A}{\d\varphi^2}\, .
		\end{equation}
		Similarly to the decomposition introduced in Eq.~(\ref{eq_couplage_premier}), we can decompose the quadratic coupling function in a universal part $\beta_u$ and a non-universal part $\bar \beta_A$. It reads
		\begin{equation}
			\beta_A(\varphi)=\beta_u(\varphi)+\bar{\beta}_A(\varphi)\, , \label{eq_couplage_second}
		\end{equation}
		where
		\begin{equation}
			\beta_u(\varphi)=\alpha_u'(\varphi) \, ,
		\end{equation}
		and
		\begin{align}
			\bar{\beta}_A(\varphi)&=(D_{\hat{m}}''(\varphi)-D_g''(\varphi))Q_{\hat{m}}^A+(D_{\delta m}''(\varphi)-D_g''(\varphi))Q_{\delta m}^A\nonumber\\
				&\quad+(D_{m_e}''(\varphi)-D_g''(\varphi))Q_{m_e}^A+D_e''(\varphi)Q_e^A \, .\label{eq_couplnonuniv2}
		\end{align}
		\begin{table*}
			\caption{Dilatonic charges of some atoms computed with Eq. \eqref{eq_qhm}, \eqref{eq_qdm}, \eqref{eq_qme} and \eqref{eq_qe}. For SiO$_2$ dilatonic charges, we compute the average of Oxygen and Silicium, as did Damour \& Donoghue \cite{damour2010prd,damour2011cqg}}\label{tabchargdilatom}
				\begin{tabular}{lllllll}
					Atom & $\mathcal{A}$ & $\mathcal{Z}$ & $Q_m \times 10^{2}$ & $Q_{dm}$ & $Q_{me} \times 10^{4}$ & $Q_e$ \\
					\hline
					Hydrogen & $1$ & $1$ & $-5.60$ & $-1.70\times 10^{-3}$ & $5.50$ & $8.20\times 10^{-4}$ \\ 
					Helium &  $4$ &  $2$ &  $-2.27$ &  $0.00$ &  $2.75$ & $6.53\times 10^{-4}$ \\
					Oxygen &  $16$ & $8$ & $-1.45$ & $0.00$ & $2.75$ & $1.48\times 10^{-3}$ \\
					Silicium &  $28.10$ & $14$ & $-1.21$ & $6.05\times 10^{-6}$ & $2.74$ & $0.205$ \\
					Iron & $56.00$ & $26.00$ & $-0.994$ & $1.21\times 10^{-4}$ & $2.55$ & $2.72\times 10^{-3}$ \\
					Magnesium & $24.30$ & $12.00$ & $-1.27$ & $2.10\times 10^{-5}$ & $2.72$ & $1.85\times 10^{-3}$ \\
					SiO$_2$ & & &             $-1.33$   & $3.02\times 10^{-6}$ & $2.75$ &  $1.76\times 10^{-3}$
				\end{tabular}
		\end{table*}

    \subsection{Modified Einstein-Infeld-Hoffmann-Droste-Lorentz equations of motion}
        In Appendix \ref{ap_pncalcul}, we present the derivation of the post-Newtonian equations of motion for $N$ massive test particles from the fields equations \eqref{eqpremierjeueqf} and \eqref{eqchampscaldil}, and the description of matter presented in Sec. \ref{subsec_intmatdil}. We show that after rescaling the constants as follow:
        \begin{subequations}\label{eqs:derived_params}
        \begin{align}
            \alpha_0&=\alpha^*_u(\varphi_0)=\sqrt{\frac{2}{Z(\varphi_0)}}\(\alpha_u(\varphi_0)-\frac{f'(\varphi_0)}{2f(\varphi_0)}\), \\ 
            \beta_0&=\beta^*_u(\varphi_0)=\frac{2}{Z(\varphi_0)}\(\beta_u(\varphi_0)+\frac{1}{2}\(\frac{f'(\varphi_0)}{f(\varphi_0)}\)^2-\frac{1}{2}\frac{f''(\varphi_0)}{f(\varphi_0)}\)-\frac{Z'(\varphi_0)}{Z(\varphi_0)^2}\(\alpha_u(\varphi_0)-\frac{f'(\varphi_0)}{2f(\varphi_0)}\) \, , \label{eq:alpha_0}\\
            \tilde{\alpha}_A& =\sqrt{\frac{2}{Z(\varphi_0)}}\bar{\alpha}_A(\varphi_0) \, , \label{eq:tilde_alpha}\\
            \tilde{\beta}_A&=\frac{2}{Z(\varphi_0)}\bar{\beta}_A(\varphi_0)-\frac{Z'(\varphi_0)}{Z^2(\varphi_0)}\bar{\alpha}_A(\varphi_0) \, , \label{eq:tilde_beta}\\
            d\beta_A&=\frac{\tilde{\beta}_A}{2}\frac{\alpha_0^2}{(1+\alpha_0^2)^2} \, , \\
            \gamma&=\frac{1-\alpha_0^2}{1+\alpha_0^2},\quad \beta=1+\frac{\beta_0}{2}\frac{\alpha_0^2}{(1+\alpha_0^2)^2}  \, , \\
            \delta_A &=\frac{\alpha_0\tilde{\alpha}_A}{1+\alpha_0^2},\quad \delta_{AB}=\frac{\tilde{\alpha}_A\tilde{\alpha}_B}{1+\alpha_0^2}  \, , \\
            \mu_A&=\frac{G}{f_0}(1+\alpha_0^2)(1+\delta_A)m_A(\varphi_0),
        \end{align}
        \end{subequations}
		where $Z(\varphi)$ is defined by Eq.~(\ref{eq_changement_scalaire}) and where the $\alpha$ and $\beta$ functions are defined in the previous section,
        the equations of motion read, at the first post-Newtonian order:
		\begin{widetext}
		\begin{align}
			\bm {a}_T=&-\sum_{A\neq T} \frac{\mu_A}{r_{AT}^3}\bm r_{AT}\left(1+\delta_T+\delta_{AT}\right)  \nonumber\\
			&-\sum_{A\neq T} \frac{\mu_A}{r_{AT}^3c^2}\bm r_{AT}\Bigg\{\gamma v_T^2+(\gamma+1)v_A^2-2(1+\gamma)\bm v_A.\bm v_T  -\frac{3}{2}\left(\frac{\bm r_{AT}.\bm v_A}{r_{AT}}\right)^2-\frac{1}{2}\bm r_{AT}.\bm a_A \nonumber \\
			&\hspace{3cm}-2(\gamma+\beta+\mathrm{d}\beta_T)\sum_{B\neq T}\frac{\mu_B}{r_{TB}}-(2\beta+2\mathrm{d}\beta_A-1)\sum_{B\neq A}\frac{\mu_B}{r_{AB}}\Bigg\} \nonumber \\
			&+\sum_{A\neq T}\frac{\mu_A}{c^2r_{AT}^3}\left[2(1+\gamma)\bm r_{AT}.\bm v_T-(1+2\gamma)\bm r_{AT}.\bm v_A\right](\bm v_T-\bm v_A) + \frac{3+4 \gamma}{2}\sum_{A\neq T} \frac{\mu_A}{c^2r_{AT}}\bm a_A \, ,\label{eq_eihmod}
		\end{align}
		\end{widetext}
		where $\mu_A$ is the gravitational parameter of body $A$, $\bm r_{AT}$ is the relative position of body $T$ with respect to $A$, $r_{AT}=\left|\bm r_{AT}\right|$ and $\bm v_A$ is the coordinate velocity of body $A$ while $\bm a_A$ is its coordinate acceleration.
		These are the modified Einstein-Infeld-Hoffmann-Droste-Lorentz (EIHDL) equations of motion.
		
		In Einstein theory of GR (that is to say without dilaton: $\gamma-1=\beta-1=\delta_T=\delta_{AT}=\mathrm{d}\beta_A=0$), these equations of motion have been first written by Lorentz \& Droste in 1917 (\cite{ld19171,ld19172}, for an English translation: \cite{lorentz1937}), then by Einstein, Infeld \& Hoffmann in 1939 \citep{eih1938}. This name composed of five personalities (EIHDL) tells better science history than only the three first names \cite{damour1990prd}.
		In appendix \ref{ap_lagr}, we give some more considerations about the dynamical system of $N$ mass monopoles in the massless dilaton theory: global Lagrangian and Hamiltonian formulation and first integrals are derived.

	\subsection{Nordtvedt effect}
		So far we have only considered test particles and have neglected their self-gravitation. In Einstein's GR, it is possible to proceed like this by virtue of the strong equivalence principle. However, in any tensor-scalar theory, this principle is broken. 
		The calculation of the strong equivalence principle violation was first done by Nordtvedt (1968) by considering extended bodies as a set of only gravitationally interacting points, and the auto gravitation energy was integrated on this set in a formalism in which the strong equivalence principle is broken \cite{nordtvedt1968pr1,nordtvedt1968pr2}. Later (1981), Will generalized this approach by modelling the bodies as perfect fluid \cite{will2018book}\footnote{The cited book was published in 2018 but the first edition was published in 1981.}. More recently (2000), Klioner \& Soffel have generalized this formalism by modelling the bodies as multipolar moments in the parametrized post-Newtonian formalism \cite{klioner2000prd}.

		A heuristic argument allows us to implement the Nordtvedt effect without performing all these integrations \cite{alsing:2012qy}. In Appendix \ref{app_nor}, we show that Nordtvedt effect can be integrated in a massless dilaton theory by substituting $\delta_A$ of EILDH equations of motion by $\delta_A'$ where
		\begin{equation}
		    \delta_A'=\delta_A-(4\beta-\gamma-3)\frac{3\mu_A}{5R_Ac^2}\label{eq_noreffect}
		\end{equation}
		where $R_A$ is the average radius of body $A$. The quantity $\mu_A/R_Ac^2$ corresponds to the self-gravitating energy of body $A$. We will use this term in all our modelling of the Solar system in the following.

	\subsection{Modified time travel}
	In addition to impacting the trajectory of planets, the dilaton theory will also impact the light propagation.
	    In tensor-scalar theory, it is known that the behaviour of light in the geometric optic approximation does not depend on the frame chosen \cite{minazzoli2014prd} (Einstein versus Jordan frames). We can then use the solutions of the field equations in the Einstein frame presented in Appendix~\ref{app:metric} and the fact light  follows the null-geodetic curves. With these approximation, a classical calculation leads to the modified time-travel between an emission event $e$ and a reception event $r$
		\begin{equation}\label{eq:shapiro}
			c(t_r-t_e)=\frac{R}{c}+\sum_A(\gamma+1-\delta_A)\frac{\mu_A}{c^2}\ln\frac{\bm{n}\cdot\bm{r}_{rA}+r_{rA}}{\bm{n}\cdot\bm{r}_{eA}+r_{eA}} \, ,
		\end{equation}
		where the $\delta_A$ parameter is given by Eq.~(\ref{eq_delta_a}), $\bm{n}=(\bm{r}_r-\bm{r}_e)/\|\bm{r}_r-\bm{r}_e\|$, $\bm{r}_{iA}=\bm{r}_i-\bm{z}_A$ and $r_{iA}=\|\bm{r}_{iA}\|$ with $i=e$ or $r$.

\section{Numerical methods with INPOP}
\label{sec_num}
We included the modifications to the equations of motion presented in Eq.~(\ref{eq_eihmod}) and to the light propagation presented in Eq.~(\ref{eq:shapiro}) in the INPOP planetary ephemerides to search for a possible signature induced by a massless dilaton. 

	\subsection{Reduction of the number of parameters}
		Constraining the WEP at the Solar system scale using a totally general phenomenological approach is difficult because, without any physical hypothesis or without considering a specific underlying theory, there are too many parameters to be constrained -- at least as many as the number of planets. For example, this would be the case if we tested the EP by including one parameter $\delta=(m_g/m_I)-1$ for each body ($m_G$ and $m_I$ being respectively the gravitational and the inertial masses). On the other hand, considering a specific theory like the massless dilaton allows one to search for a specific violation of the WEP limiting the parameters space to be explored. The parameters that characterize the theory described by the action from Eq.~(\ref{eq:action}) are the function $f(\varphi), \omega(\varphi)$ and the coupling functions characterizing the interaction between the scalar field and matter $D_i(\phi)$. At post-Newtonian level, only the background values for the function $Z(\varphi)$ defined in Eq.~(\ref{eq_changement_scalaire}) and of the background values for the first and second derivatives of the coupling functions impact the measurements.

		In the case of a linear coupling between the scalar field and matter, only the following coefficients enter the expression of the equations of motion and of the Shapiro time delay: $\gamma=\frac{1-\alpha_0^2}{1+\alpha_0^2}$, $\delta_A=\frac{\alpha_0\tilde{\alpha}_A}{1+\alpha_0^2}+(\gamma-1)\frac{|\Omega_A|}{{m}_Ac^2}$ and $\delta_{AB}=\tilde{\alpha}_A\tilde{\alpha}_B/(1+\alpha_0^2)$.  These effective parameters are related to the fundamental parameters of the theory through $\alpha_0$ which is defined by Eq.~(\ref{eq:alpha_0}) and through $\tilde \alpha_A=d_{\hat{m}}Q_{\hat{m}}^A+d_{\delta m}Q_{\delta m}^A+d_{m_e}Q_{m_e}^A+d_eQ_e^A$ (see Eqs.~\eqref{eq:tilde_alpha} and \eqref{eq_couplnonuniv}) with 
		\begin{subequations}
		\begin{align}
			d_{\hat{m}}&=\sqrt{\frac{2}{Z(\varphi_0)}}\Big[D_{\hat{m}}'(\varphi_0)-D_g'(\varphi_0)\Big]\, , \\
			d_{\delta m}&=\sqrt{\frac{2}{Z(\varphi_0)}}\Big[D_{\delta m}'(\varphi_0)-D_g'(\varphi_0)\Big]\, , \\
			d_{m_e}&=\sqrt{\frac{2}{Z(\varphi_0)}}\Big[D_{m_e}'(\varphi_0)-D_g'(\varphi_0)\Big]\, , \\
			d_e&=\sqrt{\frac{2}{Z(\varphi_0)}}[D_e'(\varphi_0)]\, .
		\end{align}
		\end{subequations}
		
		In case of a non-linear coupling, the following  coefficients enter also the modeling of the observations: $\beta=1-\beta_0\alpha_0^2/2(1-\alpha_0^2)^2$ and $d\beta_A=\tilde{\beta}_A\alpha_0^2/2(1+\alpha_0^2)^2$. These effective parameters are related to the  of the theory through $\alpha_0$, $\beta_0$ (see Eq.~\eqref{eq:alpha_0}) and  $\tilde{\beta}_A=b_{\hat{m}}Q_{\hat{m}}^A+b_{\delta m}Q_{\delta m}^A+b_{m_e}Q_{m_e}^A+b_eQ_e^A$ (see Eqs.~\eqref{eq:tilde_beta} and \eqref{eq_couplnonuniv2}) with
		\begin{subequations}
		\begin{align}
			b_{\hat{m}}&=\frac{2}{Z(\varphi_0)}\Big[D_{\hat{m}}''(\varphi_0)-D_g''(\varphi_0)\Big] - \frac{Z'(\varphi_0)}{Z(\varphi_0)}\Big[D_{\hat{m}}'(\varphi_0)-D_g'(\varphi_0)\Big] \, , \\
			b_{\delta m}&=\frac{2}{Z(\varphi_0)}\Big[D_{\delta m}''(\varphi_0)-D_g''(\varphi_0)\Big]-\frac{Z'(\varphi_0)}{Z(\varphi_0)}\Big[D_{\delta m}'(\varphi_0)-D_g'(\varphi_0)\Big] \, , \\
			b_{m_e}&=\frac{2}{Z(\varphi_0)}\Big[D_{m_e}''(\varphi_0)-D_g''(\varphi_0)\Big]-\frac{Z'(\varphi_0)}{Z(\varphi_0)}\Big[D_{m_e}'(\varphi_0)-D_g'(\varphi_0)\Big]\, , \\
			b_e&=\frac{2}{Z(\varphi_0)}D_e''(\varphi_0)-\frac{Z'(\varphi_0)}{Z(\varphi_0)}D_e'(\varphi_0)\, .
		\end{align}
		\end{subequations}
		Note that the Nordtvedt term is also modified when taking into account non-linear coupling such that $\delta_A=\frac{\alpha_0\tilde{\alpha}_A}{1+\alpha_0^2}+(\gamma+3-4\beta)\frac{|\Omega_A|}{{m}_Ac^2}$.

		
		\begin{table*}[ht]
			\caption{Dilatonic charges of the main gazeous, then telluric bodies of the Solar system. For $SiO_2$, the weighting is computed based on the number of atoms. In the last column, we compute the relative dispersion of the dilatonic charges.}
			\label{tabcharplaga}
			\begin{tabular}{llllll|l}
				               & Sun & Jupiter & Saturn & Uranus & Neptune  &\\
				\hline
				Hydrogen                    & $74\%$ & $90\%$   & $96\%$ & $83\%$ & $80\%$ &\\
				Helium                       & $25\%$ & $10\%$ & $3\%$ & $15\%$ & $19\%$  &\\
				SiO$_2$                      & $0\%$ & $0\%$ & $0\%$ & $0\%$ & $0\%$  &\\
				Iron                          & $0\%$ & $0\%$ & $0\%$ & $0\%$ & $0\%$ &\\
				Magnesium                    & $0\%$ & $0\%$ & $0\%$ & $0\%$ & $0\%$  & $\frac{\sigma_Q}{\langle Q \rangle}$ \\ 
				$\langle Q_{\hat{m}}  \rangle \times 10^{2} $ & $-4.76$ & $-5.27$ &$-5.50$&$-5.09$&$-4.96$&$ 5.6\%$ \\
				$\langle Q_{\delta m} \rangle \times 10^{3}$ &$-1.27$&$-1.53$&$-1.65$&$-1.44$&$-1.37$&$ 10\%$\\
				$\langle Q_{m_e} \rangle \times 10^{4}$ &$4.81$&$5.23$&$5.42$&$5.08$&$4.97$&$ 4.6\%$\\
				$\langle Q_e \rangle \times 10^{4}$ &$7.78$&$8.03$&$8.15$&$7.94$&$7.88$&$1.8$\%
			\end{tabular}
			
			\begin{tabular}{lllll|l}
				            & Mercury & Venus/Mars & Earth & Moon  & \\
				\hline
				Hydrogen                    & $0$\% &$0$\% &  $0$\% &$0$\% & \\
				Helium                       & $0$\% & $0$\% & $0$\% & $0$\%  & \\
				SiO$_2$                      &   $40$\% & $80$\% &$45$\%&$63$\%  & \\
				Iron                          & $60$\% & $20$\% & $32$\% & $13$\% & \\
				Magnesium                    & $0$\% & $0$\% & $14$\% & $0$\%  & $\frac{\sigma_Q}{\langle Q \rangle}$\\ 
				$\langle Q_{\hat{m}}  \rangle \times 10^{2}$&$-1.13$&$-1.26$&$-1.20$&$-1.27$&$ 5.3$\% \\
				$\langle Q_{\delta m} \rangle \times 10^{5}$     &$7.41$&$2.67$&$4.74$&$2.33 $&$ 54.5$\%\\
				$\langle Q_{m_e} \rangle \times 10^{4}$    &$2.63$&$2.71$&$2.67$&$2.71 $&$ 1.4$\% \\
				$\langle Q_e \rangle \times 10^{3}$&$2.34$&$1.95$&$2.11$&$1.93 $&$ 9.1$\%
			\end{tabular}
		\end{table*}

		In Table \ref{tabcharplaga}, we give the values of the dilatonic charges estimated for the main bodies of the Solar system, using Table \ref{tabchargdilatom}. We note that, except for $Q_{\delta m}$ in telluric bodies, all dilatonic charges have a similar value for the various telluric planets and have a similar value for the gazeous planets. More precisely, the variations of the dilatonic charges are less than 10 \% for both types of planets. Moreover, we note that in telluric bodies, $Q_{\delta m}$ is one order of magnitude smaller than the other dilatonic charges. If we assume that all coupling coefficients are of the same order of magnitude, similarly to what has been assumed in \cite{damour2010prd}, we can also safely neglect the dispersion of the value of $Q_{\delta m}$ for telluric planets. Therefore, in the following, we will consider that all telluric planets share the same dilatonic charges and that all gaseous planets share the same dilatonic charges as well. In Table \ref{tab_valuesxyzu}, we present the average values for $\langle Q_{\hat{m}} \rangle_i$, $\langle Q_{\delta m} \rangle_i$, $\langle Q_{m_e} \rangle_i$, and  $\langle Q_e \rangle_i$, for the telluric ($i=T$) and gaseous bodies ($i=G$). These assumptions allow us to reduce significantly the computational time to explore the parameters space. This hypothesis can be criticized, because, until we do not have any empirical positive detection of the coupling constant, nothing can be told about their ratio and the fact that they should have the same order of magnitude. While a more detailed modeling including more fitted parameters would be useful in case of a positive detection, the assumptions described above is sufficient for this first exploration.

       \begin{table}
           \caption{Average values of the dilatonic charges for telluric and gazeous bodies, computed from Table \ref{tabcharplaga}.}
            \label{tab_valuesxyzu}

            \centering
            \begin{tabular}{c|c|c|c|c|c|c|c|c}
            Dilatonic   & $\langle Q_{\hat{m}}\rangle_T$ & $\langle Q_{\delta m} \rangle_T$ & $\langle Q_{m_e} \rangle_T$ &  $\langle Q_e \rangle_T$ & $\langle Q_{\hat{m}} \rangle_G$ & $\langle Q_{\delta m} \rangle_G$ & $\langle Q_{m_e} \rangle_G$ &  $\langle Q_e \rangle_G$ \\    
            charge & & & & & & & & \\
            \hline
            Average & $-1.2\times 10^{-2}$ & $4.3\times 10^{-5}$ & $ 2.7\times 10^{-4}$ & $2.1\times10^{-3}$  & $-5.1\times 10^{-2}$ & $-1.5\times 10^{-3}$    & $5.1\times 10^{-4}$  & $8.0\times 10^{-4}$ \\
            value & & & & & & & & 
            \end{tabular}
        \end{table}

		Under these assumptions, the number of effective parameters impacting planetary ephemerides is reduced to 3 for the linear coupling: $\alpha_0$ given by Eq.~\eqref{eq:alpha_0}, $\tilde \alpha_T$ and $\tilde \alpha_G$ which are related to the fundamental parameters of the theory through 

		\begin{subequations}
		\begin{align}
		    \tilde{\alpha}_T &=  \sqrt{\frac{2}{Z(\varphi_0)}} \left[-1.2\times 10^{-2} \Big(D'_{\hat m}(\varphi_0) - D'_{g}(\varphi_0)\Big) +4.3\times 10^{-5} \Big(D'_{\delta m}(\varphi_0) - D'_{g}(\varphi_0)\Big)\right. \nonumber\\
		    &\phantom{========}\left.+2.7\times 10^{-4} \Big(D'_{m_e}(\varphi_0) - D'_{g}(\varphi_0)\Big)+2.1\times10^{-3}D'_{e}(\varphi_0) \right] \\
		    \tilde{\alpha}_G &= \sqrt{\frac{2}{Z(\varphi_0)}} \left[-5.1\times 10^{-2} \Big(D'_{\hat m}(\varphi_0) - D'_{g}(\varphi_0)\Big) -1.5\times 10^{-3} \Big(D'_{\delta m}(\varphi_0) - D'_{g}(\varphi_0)\Big) \right.\nonumber\\
		    &\phantom{========}\left.+5.1\times 10^{-4} \Big(D'_{m_e}(\varphi_0) - D'_{g}(\varphi_0)\Big)+8.0\times 10^{-4}D'_{e}(\varphi_0) \right]\, ,
		\end{align}
		\end{subequations}
		From now, and until the end, in order to simplify the notations, we remove the tilde for the first-order coupling parameters: we set $\alpha_A=\tilde{\alpha}_A$ for $A\in\{T,G\}$.
		If one considers non-linear couplings, three additional parameters impact planetary ephemerides: $\beta_0$, $d\beta_T$ -- which is $d\beta_A$ where $A$ runs on the telluric bodies --, and $d\beta_G$ -- which is $d\beta_A$ where $A$ runs over the gazeous bodies.

	\subsection{Introduction in planetary ephemerides}
	\label{sec:setb}
		INPOP (Int\'egration Num\'erique Plan\'etaire de l'Observatoire de Paris) is a planetary ephemeris developped since 2003 \cite{fienga2008aa}. It consists in integrating numerically the equations of motion of the main bodies and of  more than 14,000  asteroids, and adjusting the model parameters to the data with a least-squares algorithm. We model the solar system as a system of monopoles, except for the Sun, and the Earth-Moon system. For the Sun, we model its oblateness $J_{2\odot}$ defined as the dimensionless coefficient of the quadratic term of the multipole potential. For the Earth-Moon system, we model multipole interactions \cite{fienga2019inpop}. However, concerning GR and dilaton effects, only the monopoles terms are considered. Thus, for GR effects and dilaton effects, bodies are considered as homogeneous spheres. At the current level of observational accuracy, this is the level of modelisation that is necessary to minimize the residuals in the planetary ephemeris INPOP. Additional planetary parameters have been found to have no impact on the residuals. INPOP19a, includes recent data from Juno which provides accurate Jupiter barycenter positions, upgrades in the Cassini datasets which provides Saturn barycenter positions and additional Mars orbiter observations. In addition, several asteroids and a trans Neptunian objects belt have been added \cite{fienga2020MNRAS, DiRuscio2020} and we have reanalyzed Cassini data, including the recent Grande Finale \cite{DiRuscio2020}.
		We have summarized the most sensitive data in Table \ref{tab_sensdata}. The complete documentation is available in \cite{fienga2019inpop}. 

		\begin{table}[ht]
			\caption{Summary of the data sets and their average observational uncertainties $\sigma_{r}$, in meters. Messenger data where provided by \cite{verma2016jgr}. ``Cassini JPL'' data are those provided by JPL \citep{hees2014prd}. Cassini Navigation and Gravity flybys data and Grand Finale are those reduced by \cite{fienga2019inpop,DiRuscio2020}.}

		\resizebox{0.6\columnwidth}{!}{
			\begin{tabular}{cccc}
				Observations&  \# & dates & $\sigma_r$ (m) \\
				\hline
				Messenger & 1065&2011-2014 &4.1 \\
				Mars Express &  27849 &  2005-2017 &      2.0  \\
				Mars Odyssey &  18234 &  2002-2014 &      1.3  \\
				Cassini JPL &166 &2004-2014 & 25 \\
				Cassini Navigation and Gravity flybys & 614 & 2006-2016 & 6.1\\
				Cassini Grand Finale & 9 & 2017 & 2.7 \\
				Juno & 9 &  2016-2018 &      18.5 %
			\end{tabular}
		}
			\label{tab_sensdata}
		\end{table}

		INPOP is regularly used for testing fundamental physics \cite{verma2014aa, fienga2015cm, viswanathan2018mn, bernus2019prl, *bernus2020prd}. For getting a realistic constraint, we have already shown, in particular in \cite{bernus2019prl,fienga2020MNRAS,bernus2020prd} that the dilaton parameters have to be constrained together with the parameters of the reference model. The reference model contains the equations of motion and of light propagation at the first post-Newtonian approximation of GR. The method consists in adding the alternative terms in the acceleration \emph{before} the adjustment. The dilaton parameters being fixed, the parameters of the reference model are then adjusted. The adjustment explores the reference solution parameters space in the vicinity of the parameters of the reference model until the convergence is reached 
		A statistical criterion is then applied and only the dilaton parameters whose best fit satisfies this criterion are kept to build the likelihood. We repeat this operation for a large number of dilaton parameters selected randomly with a uniform distribution, and after this  we can get the final distribution. The method has been presented in \cite{fienga2020MNRAS} and relies on the computation of the adjustment likelihood (as defined in \cite{fienga2020MNRAS} or \cite{bernus2020prd} for each of the  ephemerides obtained with some randomly selected parameters and some fitted parameters).
		In the present work, we explore 3 parameters -- in the linear coupling, we have $\alpha_0$, $\alpha_T$ and $\alpha_G$. From the point of view of the numerical ressources, a Monte Carlo exploration of these three parameters is more economic than a exhaustive exploration in a three-dimensional map (while in \cite{bernus2020prd,fienga2020MNRAS}, we could do such an exhaustive 1-dimensional and 2-dimensional mapping respectively). Indeed, for one set of dilaton parameters, we need to adjust all the parameters of the reference solution several times, which means integrating the equations of motion, simulating the observations, computing the residuals, and finding the modification of the parameters with the partial derivatives, and doing it again until convergence. 
		Once this is done, we can compute the likelihood of the solution as exposed in \cite{fienga2020MNRAS,bernus2020prd}. This procedure takes more than one hour for each set of dilaton parameters. To speed up the process, we actually do parallel computations and estimate the likelihoods of several parameter sets at the same time. With our resources (see Sec. \ref{sec_ack}), we could compute 28\,800 sets of dilaton parameters. Once we have the likelihood $L(\alpha_0,\alpha_G,\alpha_T)$, as in \cite{fienga2020MNRAS,bernus2020prd} we can interpret it as the probability that the solution is better (if $L>1/2$) or worse (if $L<1/2$) than the reference solution. Here, we perform a rejection sampling (or accept-reject algorithm). This means that for each set of values of $(\alpha_0,\alpha_T,\alpha_G)$, we keep the set of values with a probability equal to $L(\alpha_0,\alpha_T,\alpha_G)$. If the ephemeris tested has lower or equal residuals than the reference ephemeris, it has more chances to be kept, and if the residuals are higher than the reference ephemeris, it has more chances to be rejected. We repeat the operation 1000 times in order to decrease the statistical fluctuations. At the end, the survival population can be interpreted as a sampling of the posterior distribution.

		Initially, we have no idea of which initial intervals of values must be chosen for the dilaton parameters. So we had to test empirically several boundaries for the initial uniform distributions of the parameters. We modified them using the following criterions:
		\begin{itemize}
			\item If the final distribution is close to zero out of the peak, compared to the initial boundaries, or if there is almost no survivors with the rejection sampling , we choose to reduce the selection interval width. The goal of this operation is to focus on the peak to increase the accuracy of the histogram of the posterior distribution.
			\item If the final distribution contains too many elements too close of the initial boundary, or equivalently, if we cannot see any peak in the histogram, then we choose to enlarge the selection interval width. The goal of this operation is to avoid missing some part of the final distribution.
		\end{itemize}

\section{Residuals and likelihood for a linear coupling}
\label{sec_resi}
\begin{figure}
	\begin{tabular}{ccc}
		\includegraphics[scale=0.4]{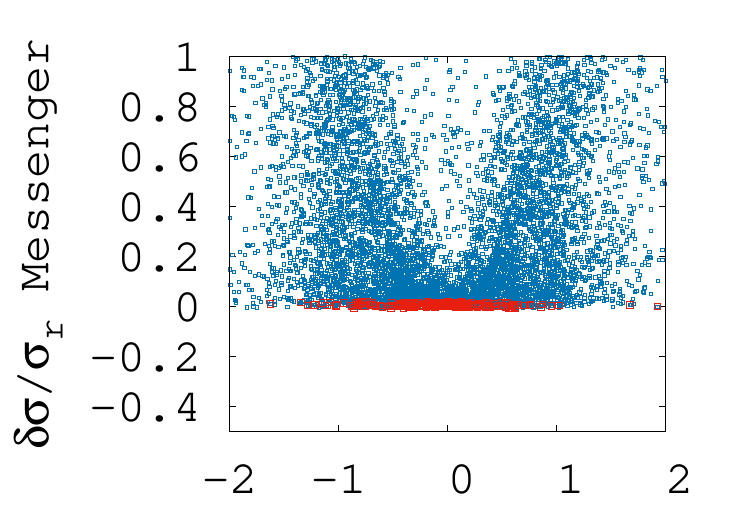} &
		\includegraphics[scale=0.4]{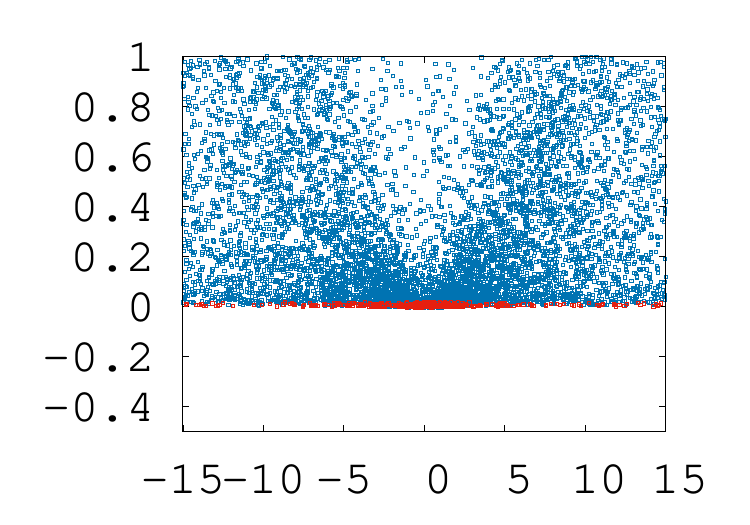} &
		\includegraphics[scale=0.4]{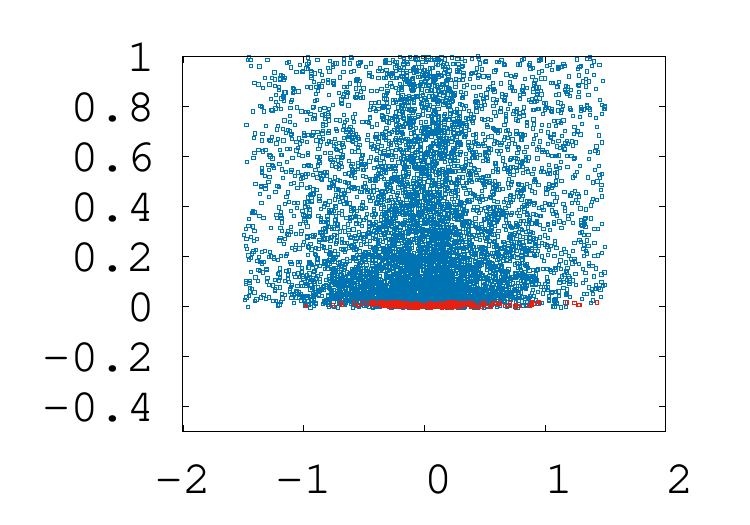} \\
		\includegraphics[scale=0.4]{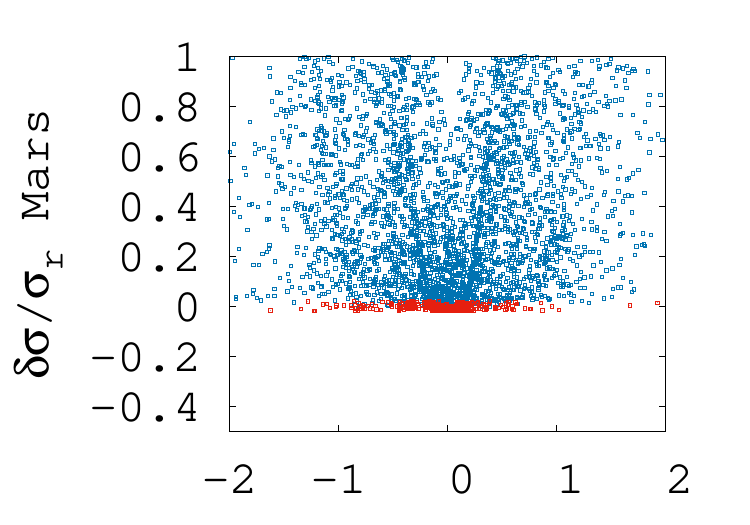} &
		\includegraphics[scale=0.4]{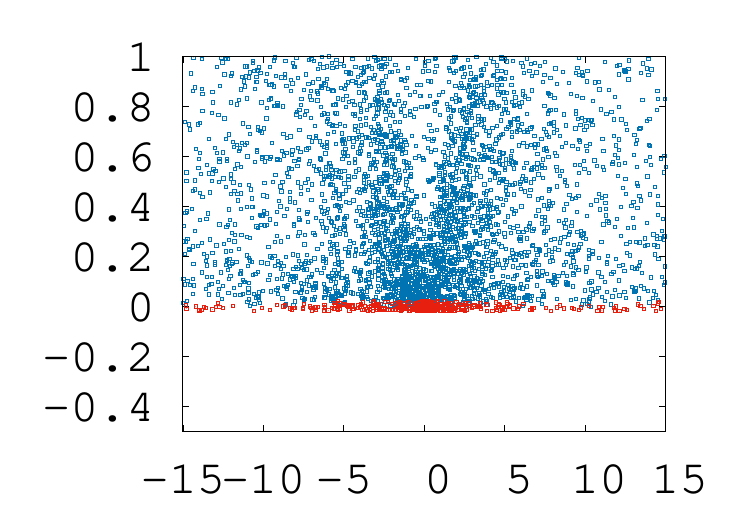} &
		\includegraphics[scale=0.4]{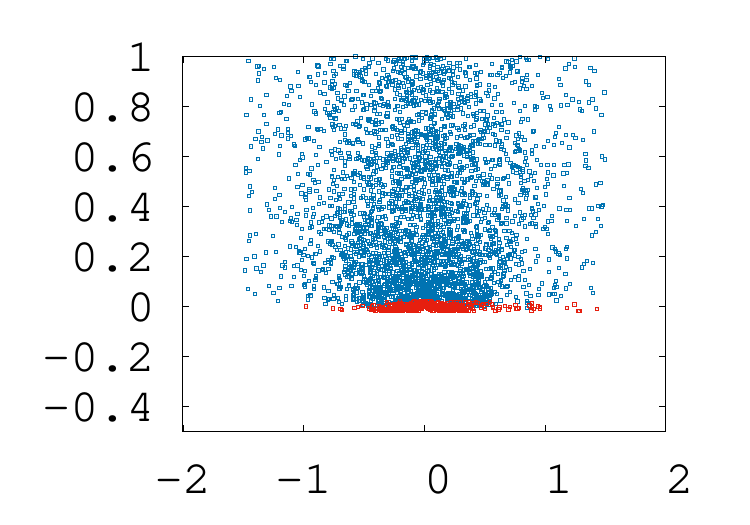} \\
		\includegraphics[scale=0.4]{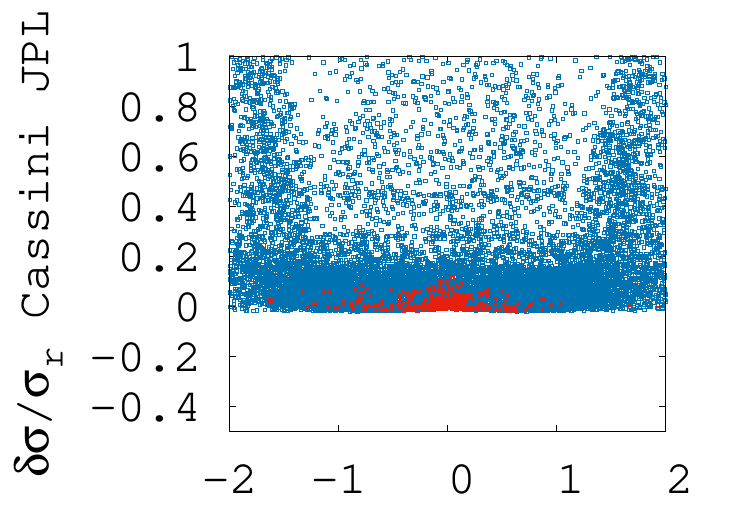} &
		\includegraphics[scale=0.4]{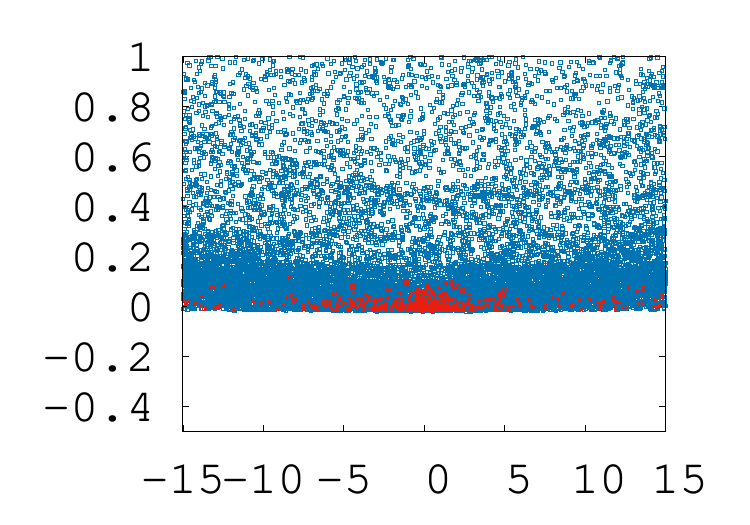} &
		\includegraphics[scale=0.4]{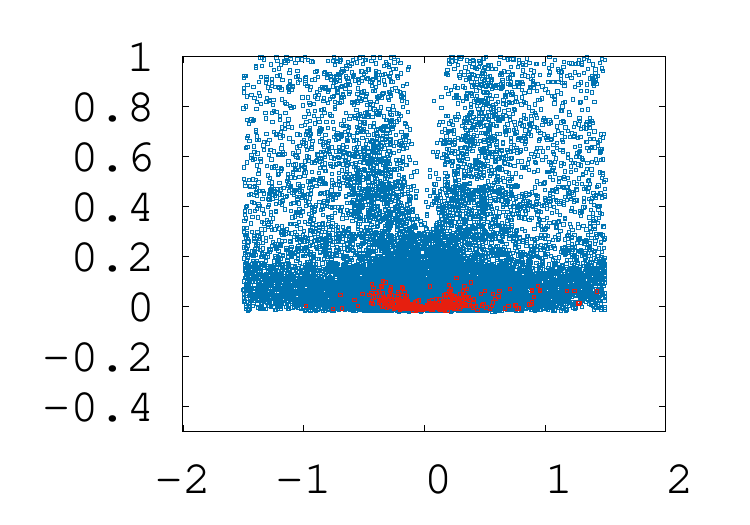} \\
		\includegraphics[scale=0.4]{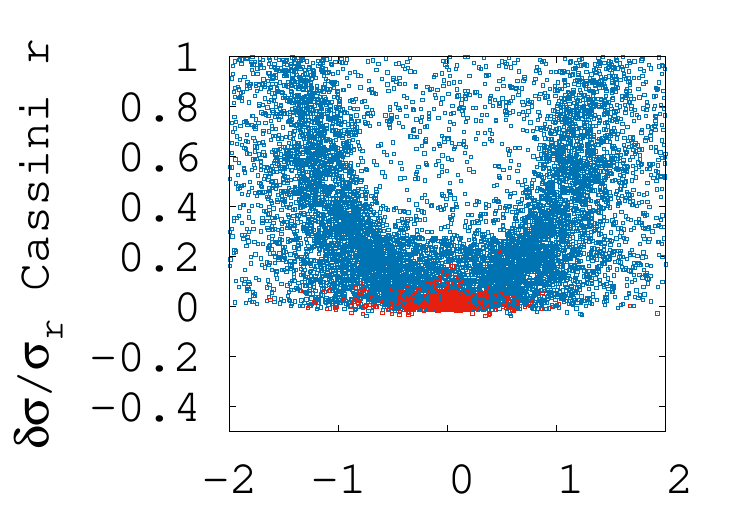} &
		\includegraphics[scale=0.4]{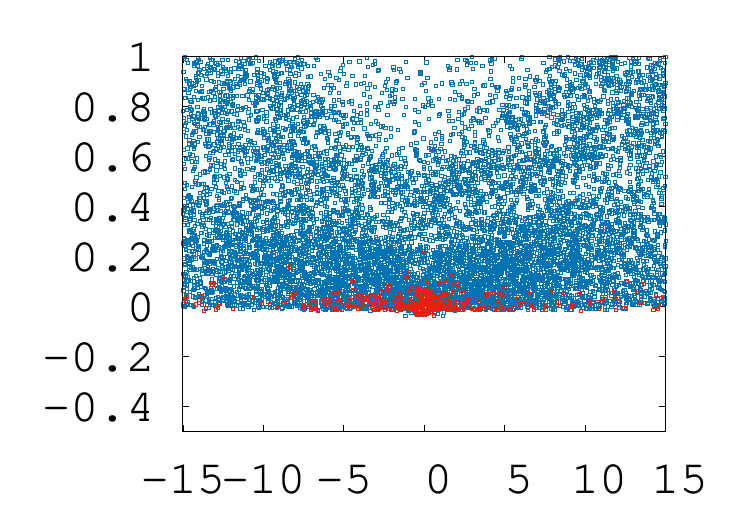} &
		\includegraphics[scale=0.4]{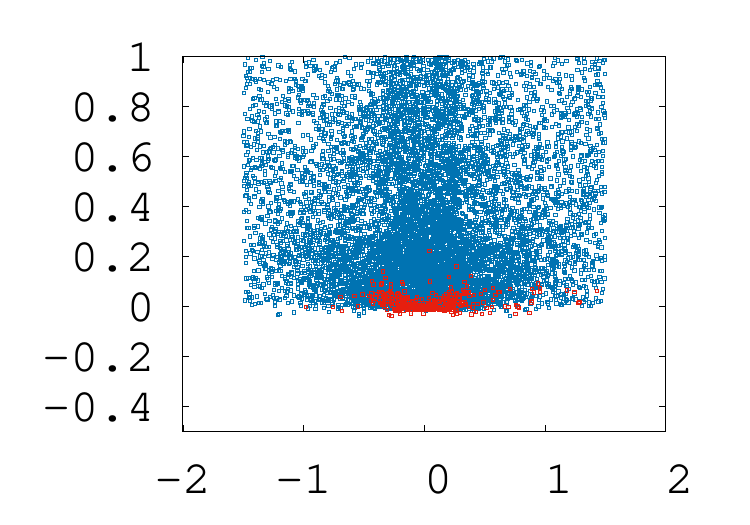} \\
		\includegraphics[scale=0.4]{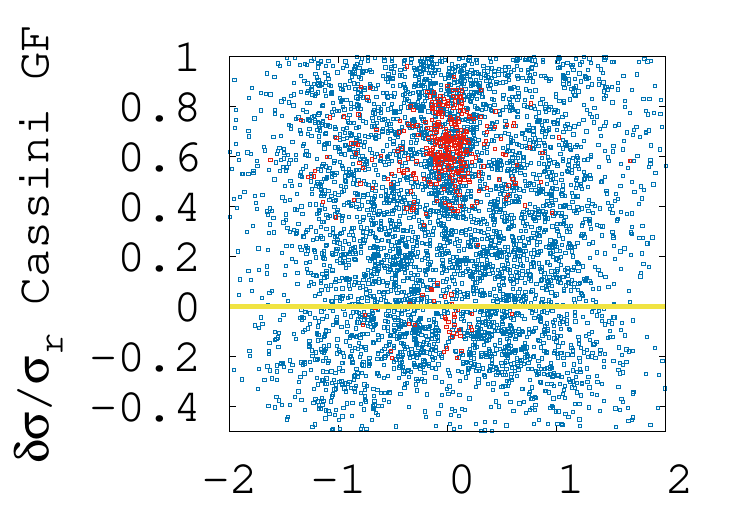} &
		\includegraphics[scale=0.4]{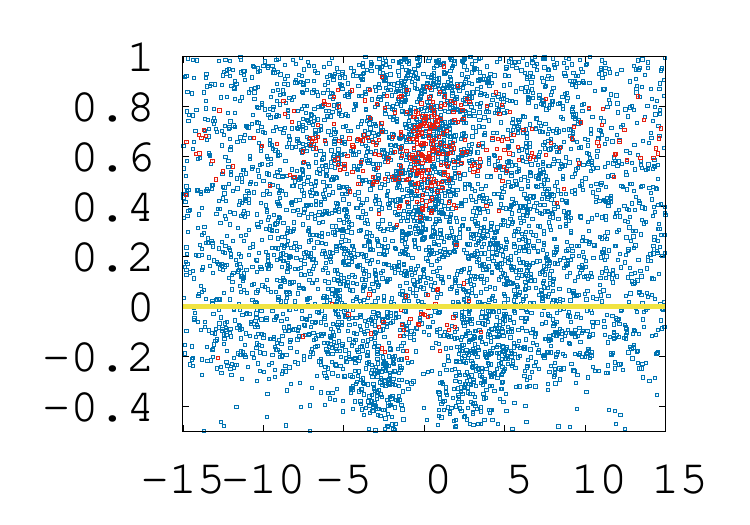} &
		\includegraphics[scale=0.4]{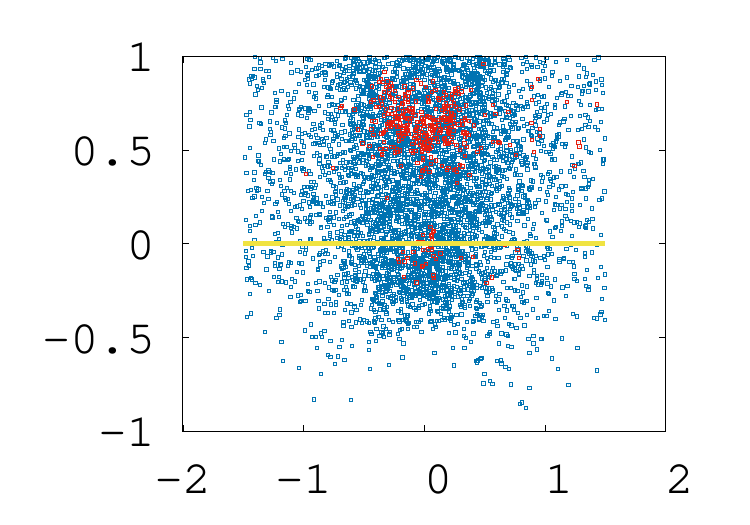} \\
		\includegraphics[scale=0.4]{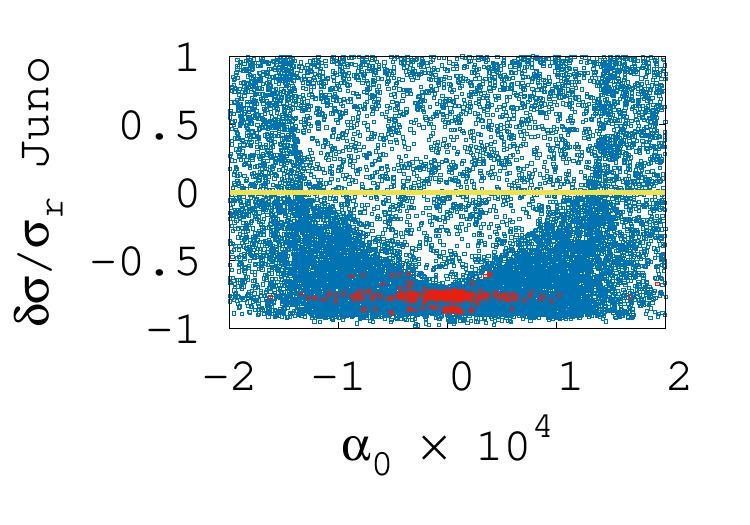} &
		\includegraphics[scale=0.4]{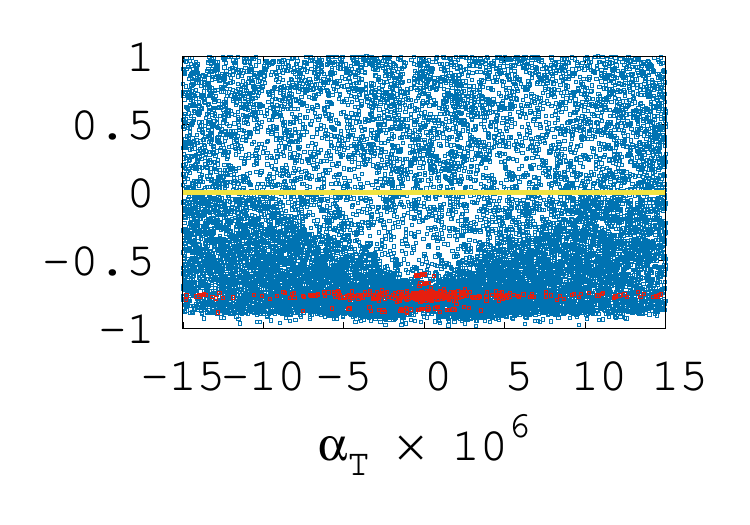} &
		\includegraphics[scale=0.4]{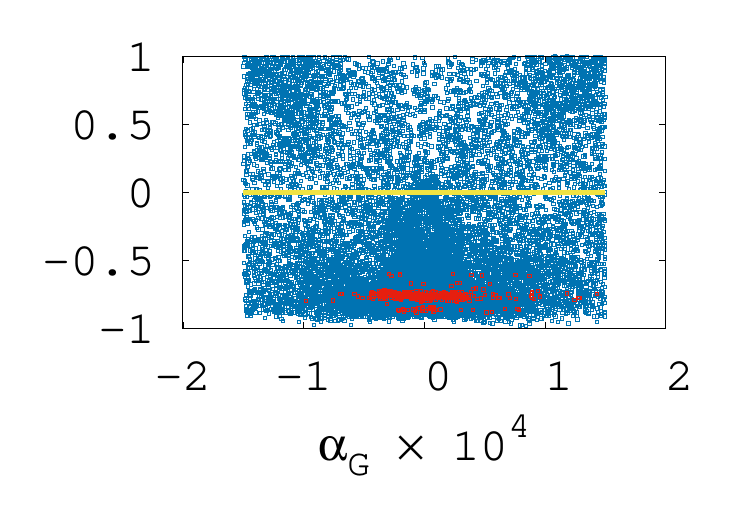}
	\end{tabular}
		\caption{Relative variations of the standard deviations ($(\sigma-\sigma_r)/\sigma_r$ where $\sigma$ is the computed standard deviation and $\sigma_r$ is the standard deviation of the reference solution) of the residuals with respect to $\alpha_0$, $\alpha_T$ and $\alpha_G$ (respectively: first column, second column, and third column). We plot in red the residuals for which $L>0.01$, that is to say the residuals of the ephemeris which have less than 99\% chances to be rejected by the algorithm. Since some residuals are better than those of the reference solution, we plot a yellow line where they are equal.}
		\label{fig_resf}
	\end{figure}

\begin{figure}
	\begin{tabular}{ccc}
		\includegraphics[scale=0.4]{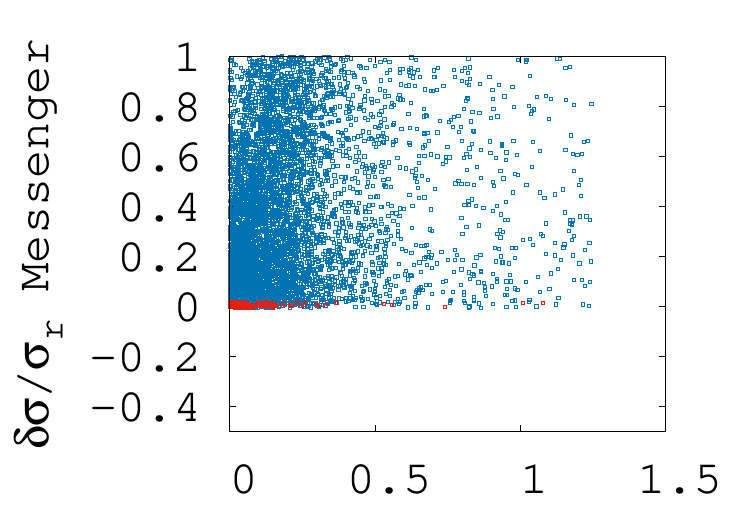} &
		\includegraphics[scale=0.4]{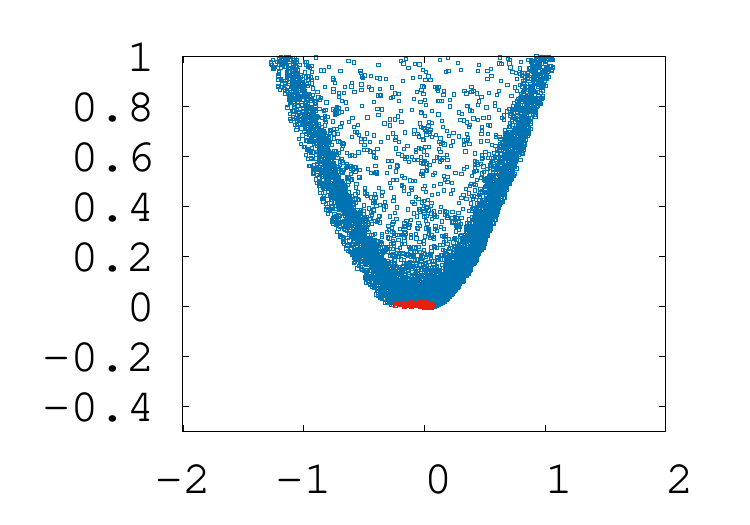} &
		\includegraphics[scale=0.4]{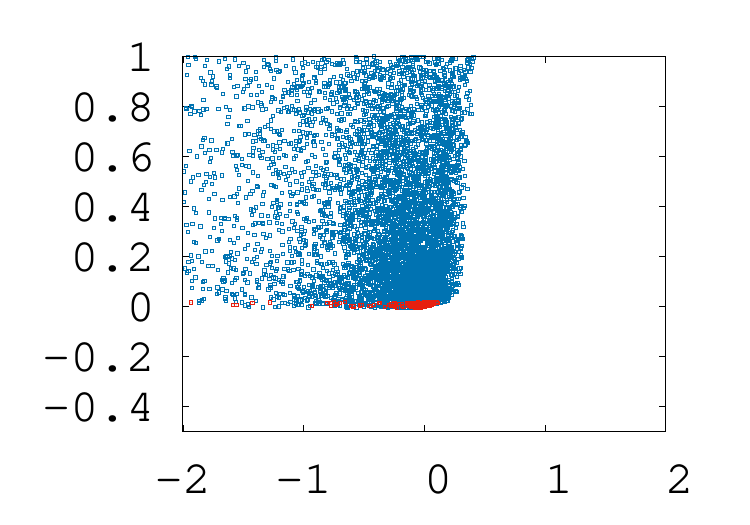} \\
		\includegraphics[scale=0.4]{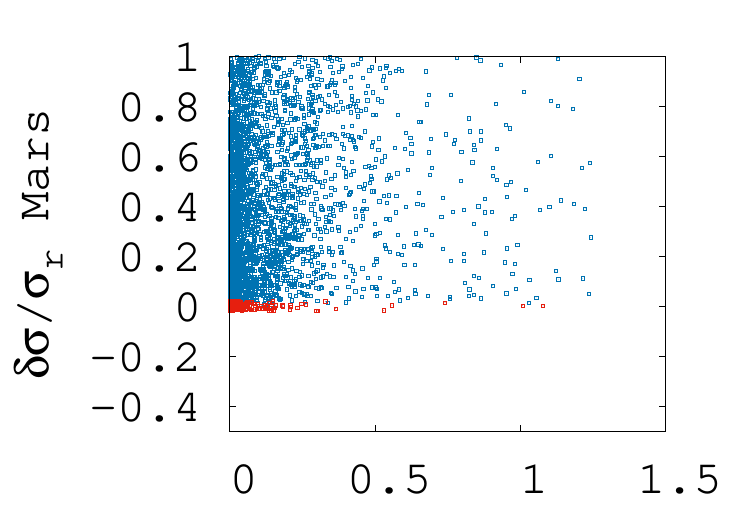} &
		\includegraphics[scale=0.4]{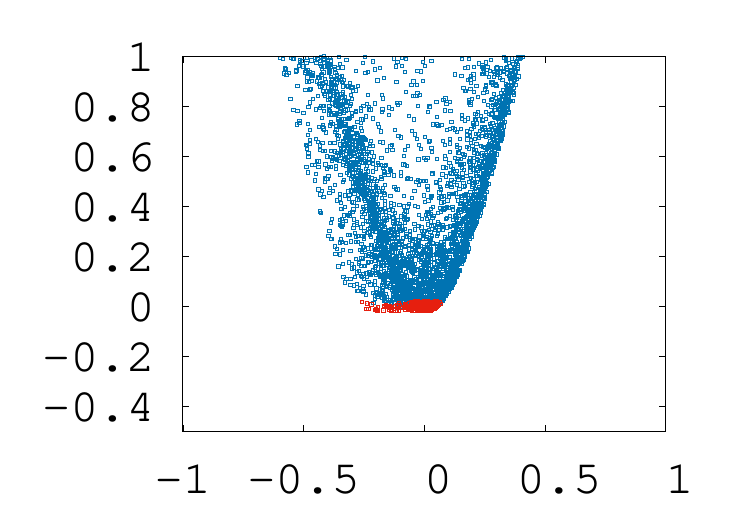} &
		\includegraphics[scale=0.4]{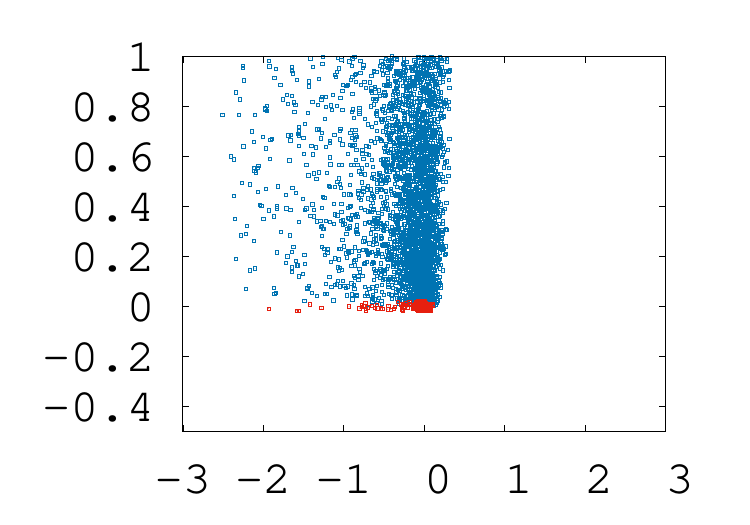} \\
		\includegraphics[scale=0.4]{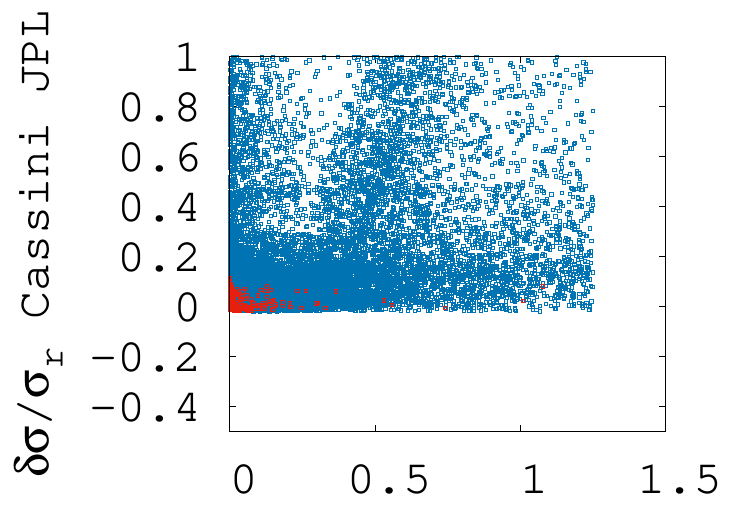} &
		\includegraphics[scale=0.4]{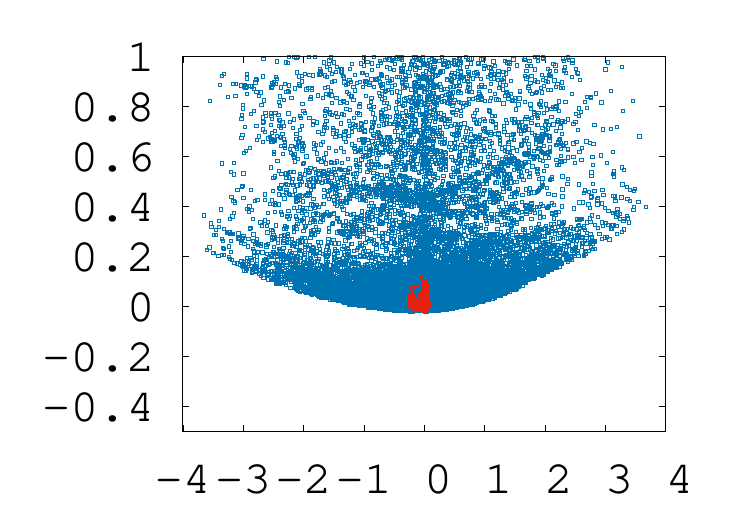} &
		\includegraphics[scale=0.4]{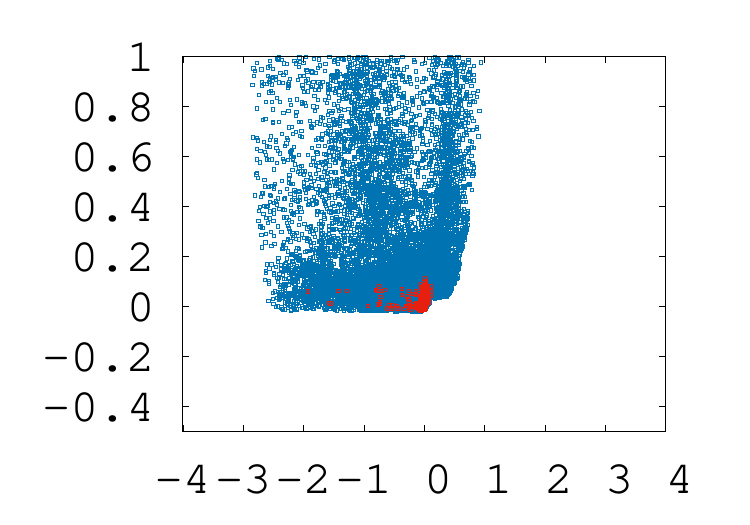} \\
		\includegraphics[scale=0.4]{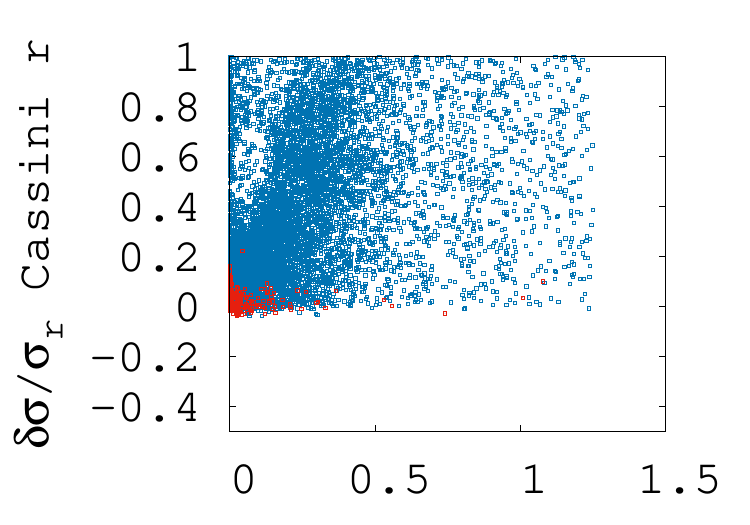} &
		\includegraphics[scale=0.4]{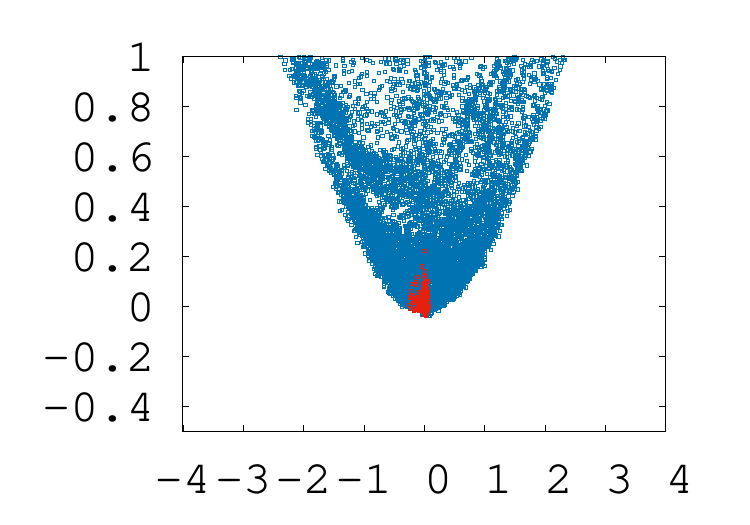} &
		\includegraphics[scale=0.4]{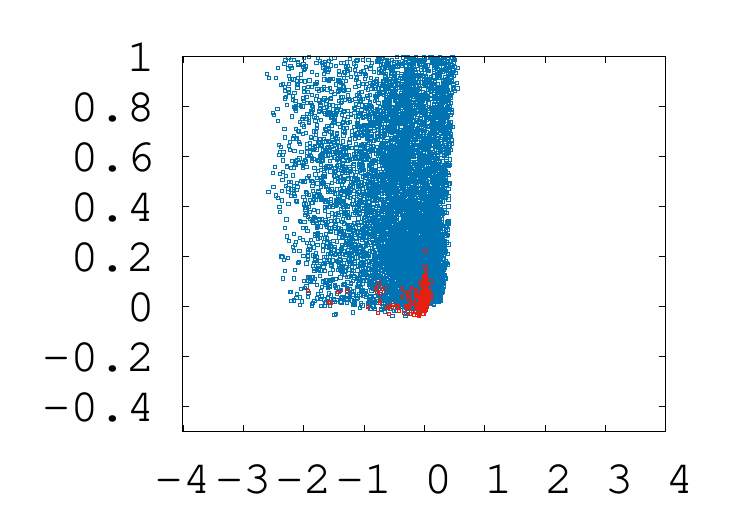} \\
		\includegraphics[scale=0.4]{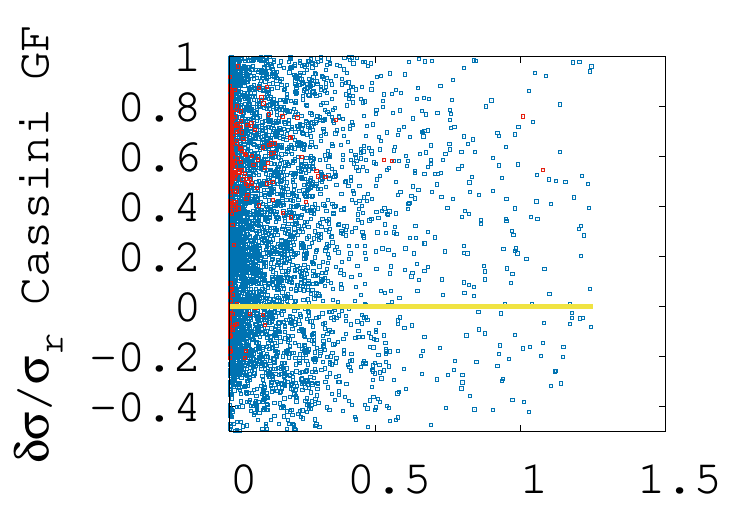} &
		\includegraphics[scale=0.4]{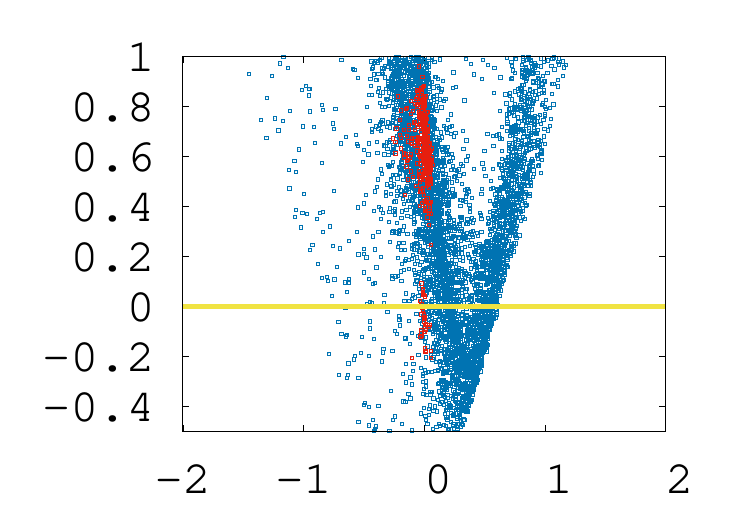} &
		\includegraphics[scale=0.4]{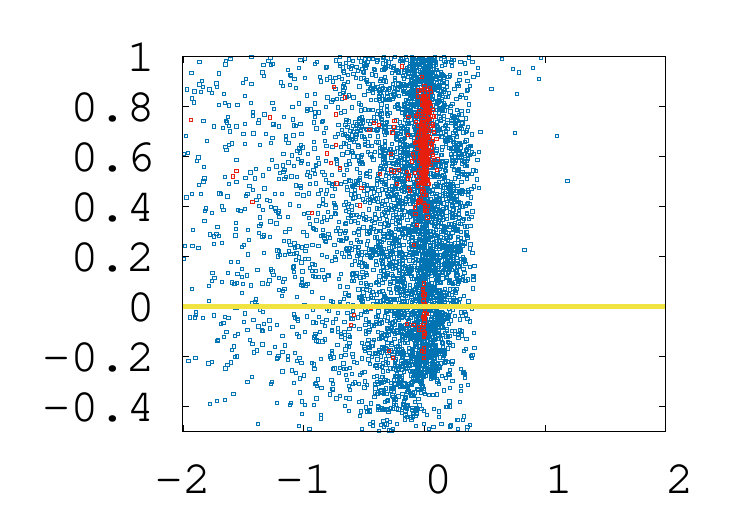} \\
		\includegraphics[scale=0.4]{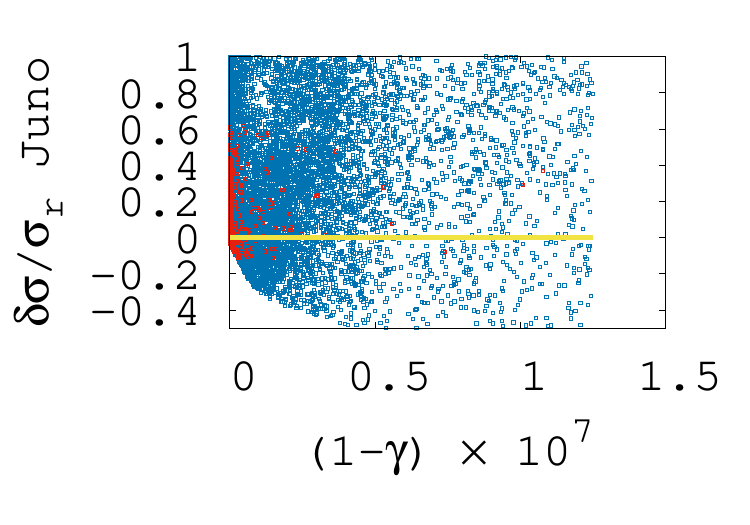} &
		\includegraphics[scale=0.4]{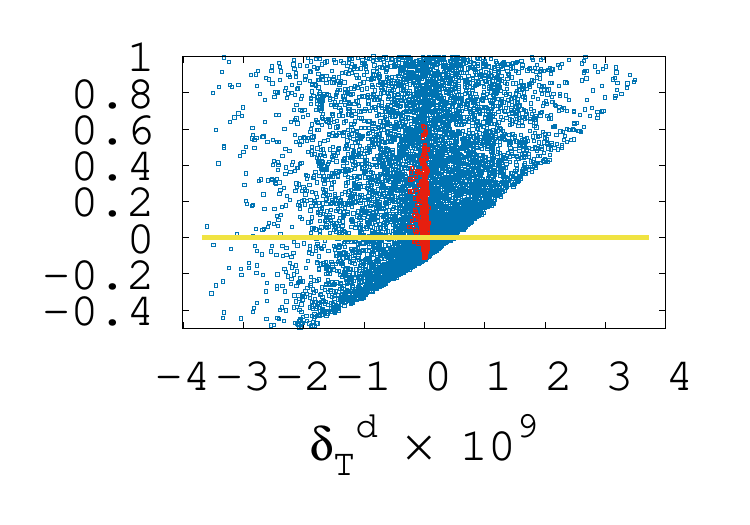} &
		\includegraphics[scale=0.4]{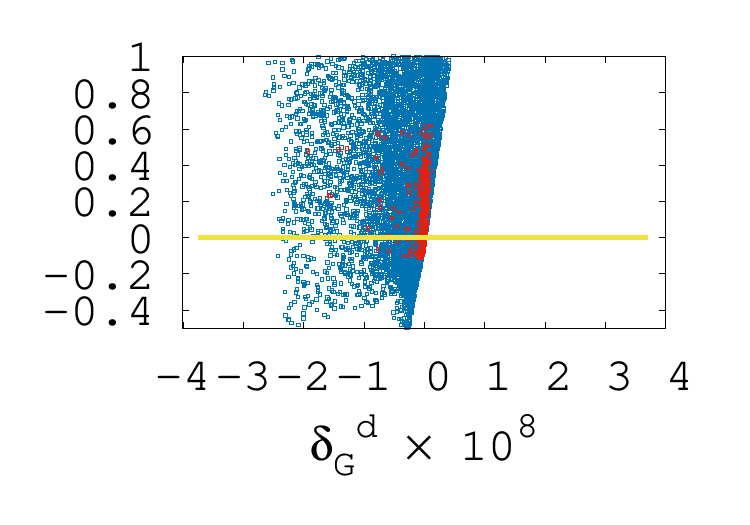}
	\end{tabular}
		\caption{Relative variations of the standard deviations ($(\sigma-\sigma_r)/\sigma_r$ where $\sigma$ is the computed standard deviation and $\sigma_r$ is the standard deviation of the reference solution) of the residuals with respect to $\gamma-1$, $\delta_T^d$ and $\delta_G^d$ (respectively: first column, second column, and third column). We plot in red the residuals for which $L>0.01$, that is to say the residuals of the ephemeris which have less than 99\% chances to be rejected by the algorithm. Since some residuals are better than those of the reference solution, we plot a yellow line where they are equal.}
		\label{fig_resGF}
	\end{figure}

	We plot the residuals with respect to $\alpha_0$, $\alpha_T$, $\alpha_G$ in Fig. \ref{fig_resf}, after adjustment of the planetary parameters with the dilaton parameters randomly chosen from a uniform distribution, but before the rejection sampling . 
	It is also interesting to plot the residuals with respect to the derived parameters, see Fig. \ref{fig_resGF}.

			It appears that the parameters $\delta_A^d$ are the most constrained by the ephemeris. Indeed, in these residuals, we can see that these parameters are the only ones for which the residuals are always increasing when the parameters are far from zero. The smallness of $\gamma-1$ compared to previously published results (around $|\gamma-1|\lesssim 10^{-4}$ \cite{verma2014aa,fienga2015cm}) can be explained by the fact that we are considering a specific model where $\gamma$ is not independent from the other fitted parameters due to the presence of $\alpha_0$ in all the derived parameters, see Eqs.~(\ref{eqs:derived_params}). By introducing a dependency between the parameters, one reduces mechanically the variability of the parameters. This leads to a reduction of the interval of possible values for $\gamma$ in the dilaton framework as far as the INPOP likelihood is concerned.

			We can see that the relative variation of the residuals with respect to the derived parameters has not the canonical quadratic behaviour expected in order to perform a classical least square algorithm. 
			This latest statement validates our approach by considering a partially free-derivatives algorithm instead of a full direct least-square inversion with heavy correlated parameters.

\begin{figure}[H]
				\includegraphics[scale=0.5]{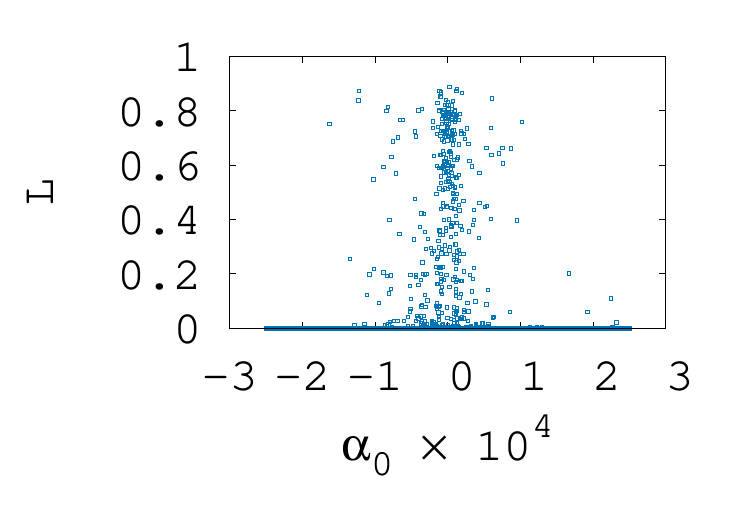}
				\includegraphics[scale=0.5]{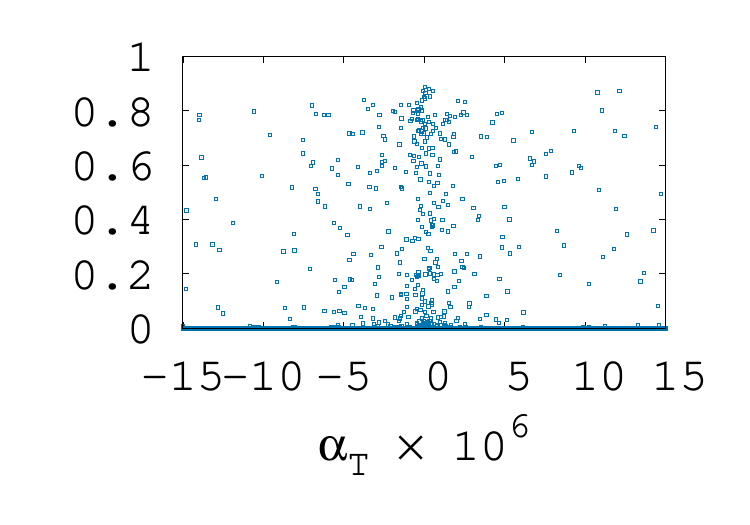}
				\includegraphics[scale=0.5]{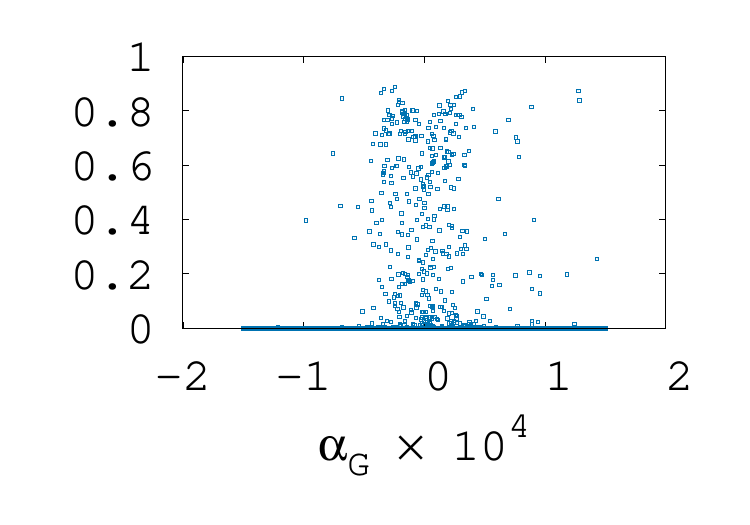}
				\caption{Likelihood with respect to $\alpha_0$, $\alpha_T$ and $\alpha_G$.}
				\label{fig_vraisf}
			\end{figure}

\begin{figure}[H]
				\includegraphics[scale=0.5]{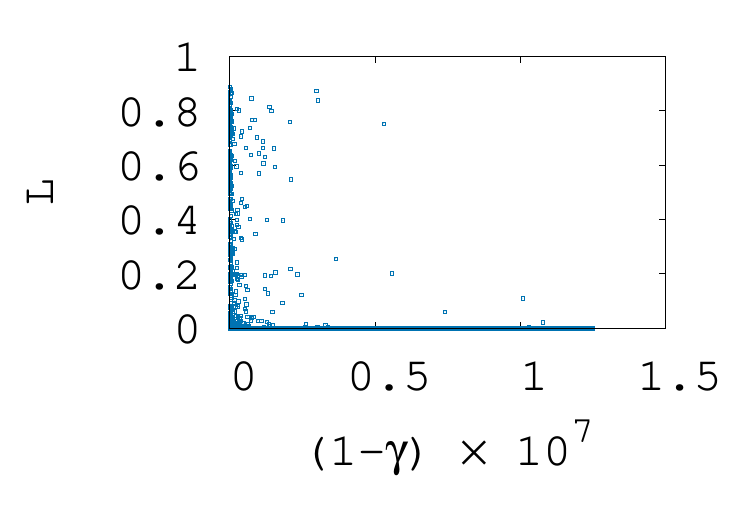}
				\includegraphics[scale=0.5]{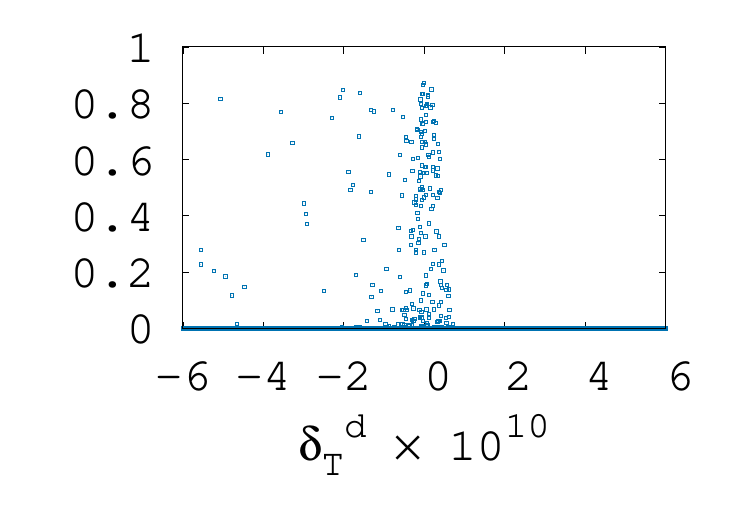}
				\includegraphics[scale=0.5]{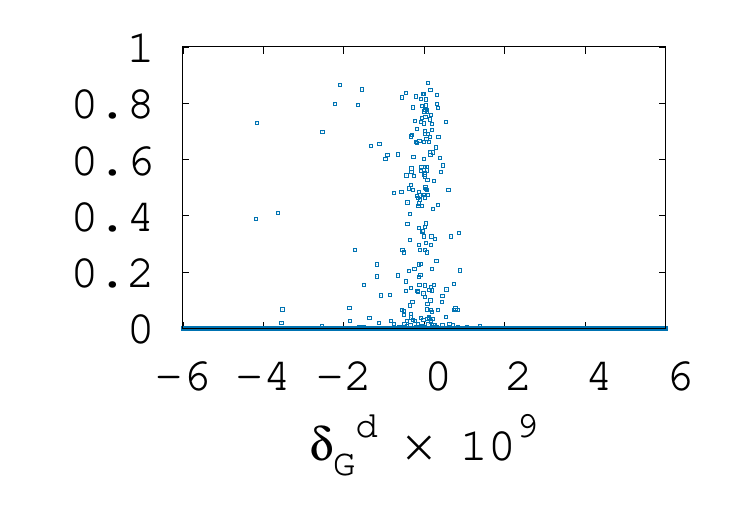}
				\caption{Likelihood with respect to $1-\gamma$, $\delta_T$ and $\delta_G$.}
				\label{fig_vraisd}
			\end{figure}

			Finally, we plot the likelihood with respect to the tested parameters (Fig. \ref{fig_vraisf}) and with respect to the derived parameters (Fig. \ref{fig_vraisd}). We get Fig. \ref{fig_vraisf} and \ref{fig_vraisd}. 
			
			We recall that for a given set of parameters tested $(\alpha_0,\alpha_G,\alpha_T)$, $L$ represents the probability to be better than the reference solution, with respect to the observational $\chi^2$ (for the reference solution itself, we have $L=1/2$). We note a blue line confounded with the abscissa axis. This shows that a lot of sets of values for $(\alpha_0,\alpha_G,\alpha_T)$ have a very low probability to be better than the reference solution. We also see that all the non zero values of $L$ are located around $0$ values for abscissa axis (which represents the tested parameters). This means that all dilaton parameters acceptable zones are compatible with zero. At this step we can already say that our results are compatible with Einstein's GR theory.

\section{Results for a linear coupling}
\label{sec_resu}
	
\begin{figure}[H]
		\includegraphics[scale=0.37]{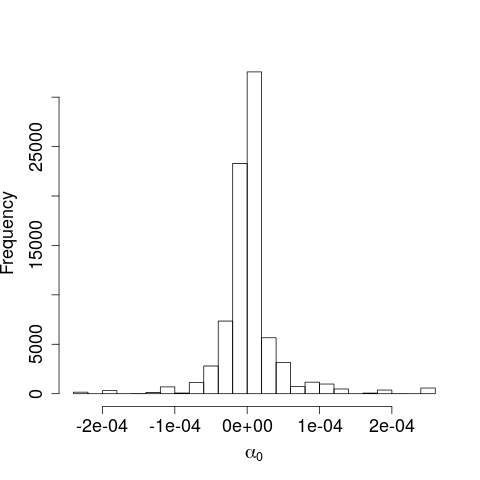}
		\includegraphics[scale=0.37]{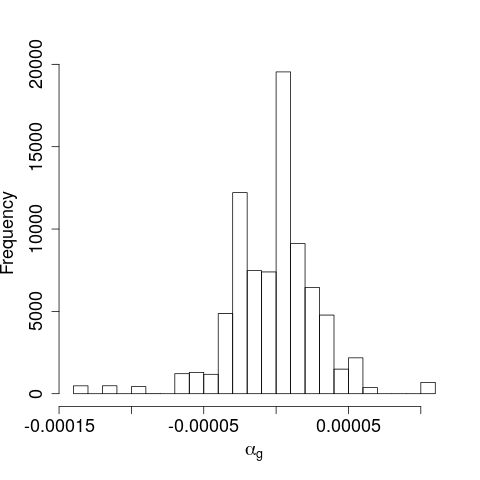}
		\includegraphics[scale=0.37]{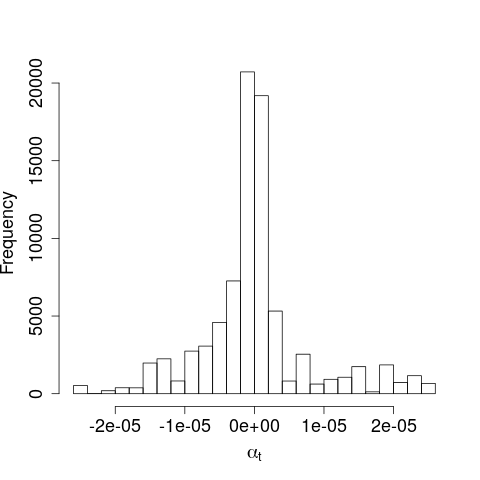}
		\caption{Histograms of dilaton parameters in the case of a linear coupling, after averaged rejection sampling .}
		\label{fig_histdillinmoy}
	\end{figure}

	After the rejection sampling, an average of $163$ solutions survive over 288\,000 solutions tested. We repeat the rejection operation 1\,000 times in order to average the statistical fluctuations and present the final results as histograms in Fig. \ref{fig_histdillinmoy}. 

\begin{figure}[H]
		\includegraphics[scale=0.37]{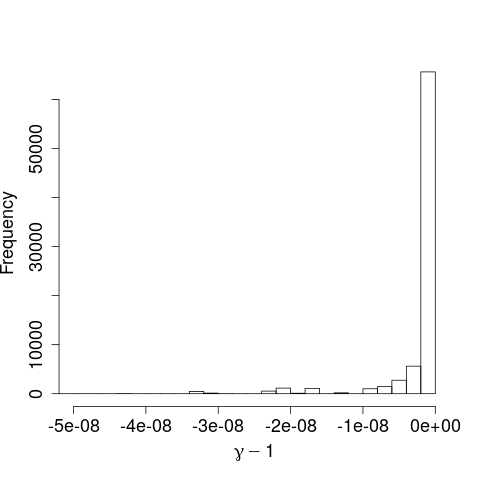}
		\includegraphics[scale=0.37]{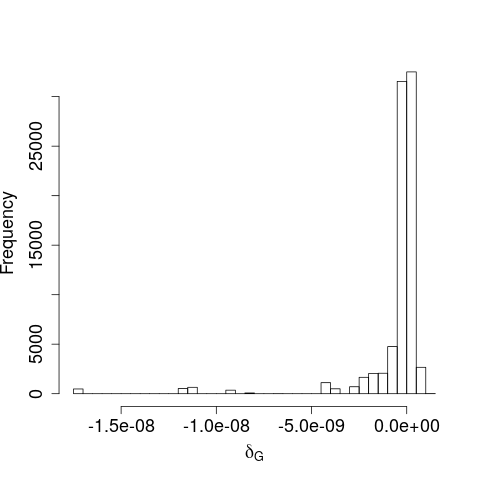}
		\includegraphics[scale=0.37]{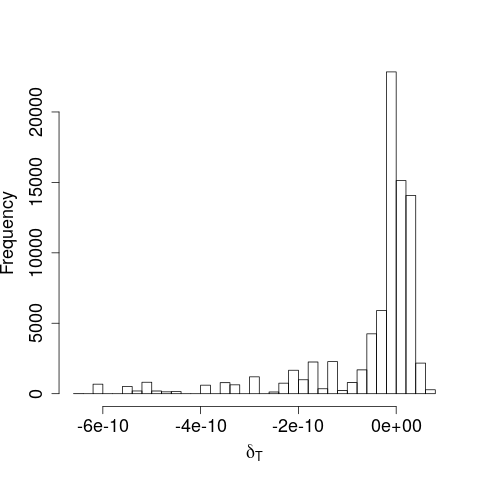}			
		\includegraphics[scale=0.37]{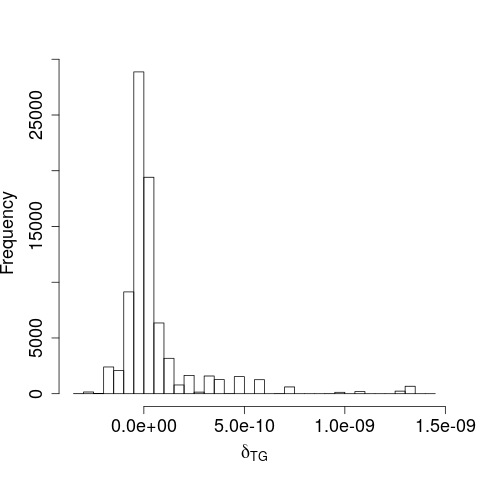}			
		\includegraphics[scale=0.37]{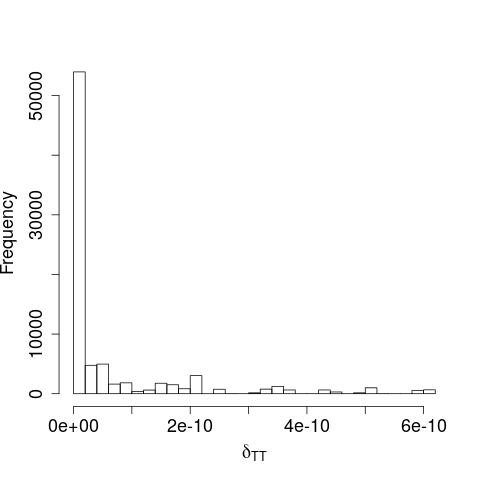}			
		\includegraphics[scale=0.37]{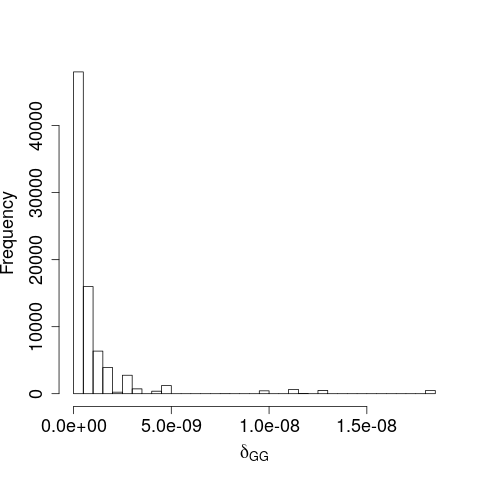}			
		\caption{Histograms of derived parameters: $\gamma-1$, $\delta_G^d$, $\delta_T^d$, $\delta_{TG}$, $\delta_{TT}$ and $\delta_{GG}$.}
		\label{fig_histderivlinmoy}
	\end{figure}

	The histograms of derived parameters (Fig. \ref{fig_histderivlinmoy} show that the  constraint on $\gamma-1$ is stronger than the usual constraints (between $10^{-4}$ and $10^{-5}$ \cite{fienga2015cm,bertotti2003nat}). As already explained, this is due to the fact that we are considering a specific model where  relation are introduced in between tested parameters with the derived parameters, which all contain $\alpha_0$.

	In Table \ref{tab_resultdillin}, we present the different quantiles associated to the different confidence levels on the tested parameters of the massless dilaton theory considering only a linear coupling between matter and the scalar field. This is for now our constraint of the massless dilaton with non-universal linear coupling. 

	\begin{table}[H]
		\begin{tabular}{lrr}
			 Confidence:      & 90\%            & 99.5\% \\
			\hline
			$\alpha_0(\times 10^{5})$  & $-0.94\pm 5.35$ & $1.01\pm23.7 $ \\
			$\alpha_T(\times 10^{6})$  & $0.24\pm1.62$     & $0.00\pm24.5 $ \\
			$\alpha_G(\times 10^{5})$  & $0.01\pm4.38$   & $-1.46\pm12.0 $
		\end{tabular}
		\caption{Final results from our analysis: confidence intervals on the linear coupling parameters for the massless dilaton theory obtained using the rejection sampling .}
		\label{tab_resultdillin}
    \end{table}

Let us note that the Lunar ephemeris alone constrains  $\Delta_{ESM} = (\delta_E-\delta_{SE}) - (\delta_M-\delta_{SM})$ to be less than $10^{-13}$ \cite{viswanathan2018mn}, where $S$, $E$ and $M$ stand for the Sun, Earth and Moon respectively. However, with the set of approximations used in the present manuscript, one has $\Delta_{ESM} =  0$ --- due to the specific linear combination that defines $\Delta_{ESM}$, and to which the Lunar ephemeris is sensitive to. What is important to note is that the planetary ephemeris, on the other hand, allow to constrain the values of $\delta_G$, $\delta_T$ and $\delta_{TG}$ individually --- that is, not a linear combination of them --- unlike the Lunar ephemeris alone, despite the greater accuracy that a Lunar ephemeris has over the whole planetary ephemeris---thanks to the much more precise data available regarding the sole position of the retro-reflectors on the Moon with respect to the Earth.
\section{Conclusion}
\label{sec_concl}
	The massless dilaton theory with non-universal coupling violates the WEP, the EEP, and the SEP. We have presented the phenomenology of this theory in the Solar system and have shown that because of the close chemical compositions of the telluric planets in one hand and of the gaseous planets in the other hand, it was possible to reduce the number of parameters to be fitted to 3 in the case of a linear coupling between the dilaton and the matter, and to 6 in the case of a quadratic coupling. In the case of a linear coupling, the parameters are $\alpha_0$, a universal coupling constant, and $\alpha_T$, $\alpha_G$, two non universal coupling constants related respectively to the telluric and gaseous bodies. In the quadratic coupling, one needs to add an additional universal constant $\beta_0$, as well as two non universal quadratic coupling constants related to the telluric bodies and gaseous bodies, $d\beta^T$ and $d\beta^G$ respectively.

	We have tested these two scenarios with the planetary ephemeris INPOP19a. To do this, we have randomly selected many values of the parameters characterizing the dilaton theory and used the method of the likelihood based on the sensitive observations in order to get the distributions of the selected parameters. We have been able to constrain the linear coupling parameters, but not the quadratic one. At 99.5\% C.L., the results read $\alpha_0=(1.01\pm23.7)\times 10^{-5}$, $\alpha_T=(0.00\pm24.5)\times 10^{-6}$, $\alpha_G=(-1.46\pm12.0)\times 10^{-5}$. 
The quadratic coupling is more difficult to constrain, due to a strong correlation between $\beta$ and $J_{2\odot}$. We have some ideas on how to solve this difficulty: namely, to push the Taylor expansion of $\beta^A_{BC}$ in the EILDH equations, and/or to use external constraint on $J_{2\odot}$ --- such as the ones from the observation of the solar oblateness \cite{meftah:2015so}, or from helioseismic data \cite{mecheri:2021mn}.

\section*{Acknowledgement}
\label{sec_ack}
	This work was granted access to the HPC resources of MesoPSL financed
	by the Region Ile de France and the project Equip@Meso (reference
	ANR-10-EQPX-29-01) of the programme Investissements d'Avenir supervised
	by the Agence Nationale pour la Recherche. It had also benefit from the support of the French Space Agency( CNES). 

\bibliographystyle{unsrt}
\bibliography{dilaton}

\appendix
\section{Post-Newtonian derivation of the equations of motion}
\label{ap_pncalcul}

	\subsection{Stress-energy tensor}
		We consider a set of test particles which do not interact. The action of these points can be expressed as follows \cite{minazzoli2016prd}
		\begin{equation}
			S_m=-c^2\sum_A\int m_A(\varphi)\ \d\tau_A=-c^2\sum_A\int m_A^*(\Phi)\ \d\tau_A^*
		\end{equation}
		where
		\begin{equation}
			m_A^*(\phi)=\sqrt{\frac{f_0}{f(\varphi)}}m_A(\varphi) \label{eq_transfertmass}
		\end{equation}
		We note that $m_A^*(\phi_0)=m_A(\varphi_0)$.
		The action can be expressed as a quadri-volumic integral
		\begin{equation}
			S_m=-\int \sum_A\rho_A^*\sqrt{-g_*}\d^4x
		\end{equation}
		where
		\begin{equation}
			\rho_A^*=\frac{c^2m_A^*(\phi(x^\mu))}{\sqrt{-g_*}u_{A*}^0}\delta^{(3)}(\bm{x}-\bm{z}_A(t))
		\end{equation}
		where $\delta^{(3)}$ is the three dimensional Dirac distribution, $z_A^\mu(t)$ are the coordinates of $A$ in the map $(x^\mu)$ and $u_{A*}^\mu=\d z_A^\mu/\sqrt{-g^*_{\alpha\beta}\d z_A^\alpha \d z_A^\beta}$ are the coordinates of the 4-velocity of $A$. From here we deduce
		\begin{equation}
			T_*^{\mu\nu}=\sum_A\rho_A{^*}u_{*A}^\mu u_{*A}^\nu
		\end{equation}
		\begin{equation}
			T_*=\sum_A\rho_{A}^*
		\end{equation}
		The mass that appears in the density is a function of $\phi$. In post-Newtonian formalism, the variation of $\phi$ is a Taylor expansion, such that we will have $\phi-\phi_0=O(c^{-2})$. Then it is also possible to express $m_*(\phi)$ as a Taylor expansion
		\begin{equation}
			\frac{m_{A}^*(\phi)}{m_{A,0}^*}=1+(\phi-\phi_0)\alpha_{A}^*(\phi_0)+\frac{1}{2}(\phi-\phi_0)^2\beta_A^*(\phi_0)+O(c^{-6})\label{eq_massphi}
		\end{equation}
		where we have, from Eq. \eqref{eq_changement_scalaire}, \eqref{eq_couplage_premier}, \eqref{eq_couplage_second} and \eqref{eq_transfertmass}
		\begin{equation}
			\alpha_A^*=\frac{\d\ln m_A^*}{\d\phi}=\alpha_u^*(\varphi)+\tilde{\alpha}_A(\varphi) 
		\end{equation}
		\begin{equation}
			\beta_A^*(\varphi)=\frac{\d\alpha_A^*}{\d\phi}=\beta_u^*(\varphi)+\tilde{\beta}_A(\varphi) 
		\end{equation}
		where
		\begin{equation}
			\alpha_u^*(\varphi)=\sqrt{\frac{2}{Z(\varphi)}}\(\alpha_u(\varphi)-\frac{f'(\varphi)}{2f(\varphi)}\) \label{eq_alphaustar}
		\end{equation}
		\begin{equation}
			\tilde{\alpha}_A=\sqrt{\frac{2}{Z(\varphi)}}\bar{\alpha}_A(\varphi) \label{eq_constcoupnonuniv}
		\end{equation}
		\begin{align}
			\beta_u^*(\phi)&=\frac{2}{Z(\varphi)}\(\beta_u(\phi)+\frac{1}{2}\(\frac{f'(\varphi)}{f(\varphi)}\)^2-\frac{1}{2}\frac{f''(\varphi)}{f(\varphi)}\)\nonumber\\
			&\quad-\frac{Z'(\varphi)}{Z(\varphi)^2}\(\alpha_u(\varphi)-\frac{f'(\varphi)}{2f(\varphi)}\)
		\end{align}
		\begin{equation}
			\tilde{\beta}_A(\phi)=\frac{2}{Z(\varphi)}\bar{\beta}_A(\varphi)-\frac{Z'(\varphi)}{Z(\varphi)^2}\bar{\alpha}_A(\varphi) 
		\end{equation}

	\subsection{Post-Newtonian solution to the field equations in the Einstein frame}\label{app:metric}
		We consider $N$ mass monopoles and their worldlines in $\M$. We define a map of $\M$, $(x^\mu)$, such that at infinity, the metric tensor is the cartesian Minkowski metric: $\|x^i\|\to\infty \Rightarrow g^*_{\mu\nu}\to\eta_{\mu\nu}=\mathrm{diag}(-1,1,1,1)$.
		In this map, the worldlines of each body are described by four functions
		\begin{equation}
			\mathcal{L}_A : t\mapsto z_A^\mu(t)
		\end{equation}
		that we can reduce to three if we consider $t$ as the time-coordinate. This choice is always possible for massive particles. We assume that the gravitational fields in which the particles are weak ($GM/cr\ll1$) and the velocities are also weak ($v/c\ll1$). In these approximations, and in Einstein frame, the consequence is
		\begin{equation}
			T_*^{00}=O(c^{+2}),\quad T_*^{0i}=O(c^{+1}),\quad T_*^{ij}=O(c^{0})
		\end{equation}
		Moreover, since we know that $\phi-\phi_0=O(c^{-2})$, then we have $(\partial_i\phi,\partial_0\phi)=O(c^{-2},c^{-3})$, such that from of Eq. \eqref{eqpremierjeueqf}, one deduces that there exist a coordinate system that satisfies the strong isotropy condition as follows \cite{DSX1991}  		\begin{equation}
			g_{00}^*=-1+2\frac{w_*}{c^2}-2\frac{w_*^2}{c^4}+O(c^{-6})
		\end{equation}
		\begin{equation}
			g^*_{0i}=-4\frac{w_*^i}{c^3}+O(c^{-5})
		\end{equation}
		\begin{equation}
			g^*_{ij}=\delta_{ij}\( 1+2\frac{w_*}{c^2} \)+O(c^{-4})
		\end{equation}
		where $w_*$ and $w_*^i$ are four fields that parametrize the metric tensor and that have to be determined.
		When linearizing Eq. \eqref{eqpremierjeueqf}, and using the time component of the harmonic gauge equation --- that is $g_*^{\alpha \beta} \Gamma^0_{* \alpha \beta} = 0$, where $\Gamma_*$ is the Christoffel symbol constructed upon the Einstein-frame metric --- we obtain partial differential equations that determine the potentials $w_*$ and $w_*^i$:
		\begin{equation}
			\Box w_*=-4\pi G\frac{T_*^{00}+\delta_{ij}T_*^{ij}}{c^2}+O(c^{-4})
		\end{equation}
		\begin{equation}
			\Delta w_*^i=-4\pi G \frac{T_*^{0i}}{c} + O(c^{-4}) 
		\end{equation}
		Linearizing Eq. \eqref{eqchampscaldil} leads to a partial derivative equation that determines the scalar field $\phi$:
		\begin{equation}
			\Box \phi = -4\pi G\frac{\partial T_*}{c^2\partial \phi}+O(c^{-6})
		\end{equation}
		From here, a straightforward perturbative calculation --- which can also be deduced from the method of Damour \& Esposito-Far\`ese \cite{damour1992cqg} --- leads to the post-Newtonian solution of the field equations
		\begin{align}
			\phi&=\phi_0-\sum_A\alpha^*_{A0}\frac{G_*m_{A0}}{c^2r_A}\[1+\frac{\bm{r}_A\cdot\bm{a}_A}{2c^2}
			             +\frac{(\bm{r}_A\cdot\bm{v}_A)^2}{2c^2r_A^2}\phantom{\sum_A}\right.\nonumber\\
			    &\left.\phantom{\sum_A}  
			             +\frac{1}{c^2}\sum_{B\ne A}\frac{G_*m_{B0}}{r_{AB}}\(1+\alpha^*_{A0}\alpha^*_{B0}+\frac{\beta^*_{B0}\alpha^*_{B0}}{\alpha^*_{A0}}\)\] 
		\end{align}
		where 
		\begin{equation}
			w_*=w_{0*}-\frac{1}{c^2}\Delta_*+O(c^{-4}) \label{eq_wet}
		\end{equation}
		\begin{equation}
			w_*^i=\sum_A\frac{G_*m_{A0}}{r_A}v_A^i+O(c^{-2}) \label{eq_wiet}	
		\end{equation}
		where
		\begin{equation}
			w_0^*=\sum_A\frac{G_*m_{A0}}{r_A}
		\end{equation}
		\begin{align}
			\Delta_*&=\sum_A\frac{G_*m_{A0}}{r_A}\[ \frac{\bm{r}_A\cdot\bm{a}_A}{2}+\frac{(\bm{r}_A\cdot\bm{v}_A)^2}{2r_A^2}-2v_A^2\phantom{\sum_A}\right.\nonumber\\
			&\quad\quad\quad\quad\quad\quad\left.+\sum_{B\ne A}(1+\alpha^*_{A0}\alpha^*_{B0})\frac{G_*m_{B0}}{r_{AB}} \]
		\end{align}
		where $\alpha_{A0}^*=\alpha_A^*(\phi_0)$, $\beta_{A0}^*=\beta_A^*(\phi_0)$, $\bm{r}_A=\bm{x}-\bm{z}_A$, $r_A=\|\bm{r}_A\|$, $\bm{a}_A=\d\bm{v}_A/\d t$.

	\subsection{Equations of motion}
	\label{sec_eqmvt}
		A way to deduce the equations of motion is to derive the Euler-Lagrange equation of a test particle in the gravitational field built by the other particle
		\begin{equation}
			L_T=-c^2\int m^*_T(\phi)\sqrt{g^*_{\mu\nu}v_T^\mu v_T^\nu}
		\end{equation}
		Performing a post-newtonian expansion using the solution of the field equations and Eq. \eqref{eq_massphi} leads to the following Lagrangian
		\begin{align}
			L_{T}&=-m_{T0}c^2+m_{T0}\frac{v_T^2}{2}+\sum_{A\ne T}\frac{G_{AT}m_{A0}m_{T0}}{r_{AT}}+\frac{m_{T0}v_{T0}^4}{8c^2} \nonumber\\
			&\quad +\frac{1}{c^2}\sum_{A\ne T}\frac{G_{AT}m_{A0}m_{T0}}{r_{AT}}\[-2(1+\gamma_{AT})\bm{v}_A\cdot\bm{v}_T\right.\nonumber\\
			&\quad
			+\frac{2\gamma_{AT}+1}{2}v_T^2 
			+(\gamma_{AT}+1)v_A^2\nonumber\\
			&\left.\phantom{\sum_A} -\frac{(\bm{v}_A\cdot\bm{r}_{AT})^2}{2r_{AT}^2}-\frac{\bm{r}_{AT}\cdot\bm{a}_A}{2}\] \nonumber\\
			&\quad -\frac{1}{c^2}\sum_{A\ne T}\frac{G_{AT}m_{A0}m_{T0}}{r_{AT}}\[ 
			                     \sum_{B\ne A}\frac{G_{AB}m_{B0}}{r_{AB}}(2\beta^A_{BT}-1) \phantom{\sum_A}\right.\nonumber\\
			&\quad\left.\phantom{\sum_A}
			                    +\sum_{B\ne T}\frac{G_{BT}m_{B0}}{r_{BT}}\(\beta^T_{AB}-\frac{1}{2}\)\]
		\end{align}
		where $\bm{r}_{AT}=\bm{x}_T-\bm{x}_A$, $r_{AT}=\|\bm{r}_{AT}\|$,
		\begin{equation}
			G_{AB}=\frac{G}{f_0}(1+\alpha_{A0}^*\alpha_{B0}^*)
		\end{equation}
		\begin{equation}
			\gamma_{AT}=\frac{1-\alpha_{A0}^*\alpha_{T0}^*}{1+\alpha_{A0}^*\alpha_{T0}^*}
		\end{equation}
		\begin{equation}
			\beta^T_{AB}=1+\frac{\beta_{T0}^*}{2}\frac{\alpha_{A0}^*}{1+\alpha_{A0}^*\alpha_{T0}^*}\frac{\alpha_{B0}^*}{1+\alpha_{B0}^*\alpha_{T0}^*} \label{eq_betaabc}
		\end{equation}
		Actually, this Lagrangian was already obtained by Damour \& Esposito-Far\`ese \cite{damour1992cqg}, but with a universal conformal coupling and not a non-universal one. However, they considered the conformal coupling in a very general way and did all the calculations with $\alpha_A=\d\ln m_A/\d\phi$ without assuming anything about the nature of the coupling during the calculations, so we can use their work for checking our equations of motion.

		Nevertheless, this Lagrangian can be simplified by taking into account the composition independent and dependent nature of some parameters. First, we can decompose the coupling constants in a universal part and a non-universal one
		\begin{equation}
			\alpha_{A0}^*=\alpha_0+\tilde{\alpha}_A
		\end{equation}
		where $\alpha_0=\alpha_u^*(\phi_0)$ (eq. \eqref{eq_alphaustar}) is the universal coupling constant and $\tilde{\alpha}_A$ has been defined in Eq. \eqref{eq_constcoupnonuniv}. We can then redefine the gravitational constant as follows
		\begin{equation}
			G_{AB}=\tilde{G}(1+\delta_A+\delta_B+\delta_{AB})
		\end{equation}
		where
		\begin{equation}
			\tilde{G}=\frac{G}{f_0}(1+\alpha_0^2)
		\end{equation}
		\begin{equation}
			\delta_A=\frac{\alpha_0\tilde{\alpha}_A}{1+\alpha_0^2} \label{eq_delta_a}
		\end{equation}
		\begin{equation}
			\delta_{AB}=\frac{\tilde{\alpha}_A\tilde{\alpha}_B}{1+\alpha_0^2} \label{eq_delta_ab}
		\end{equation}
		Most of the alternative constant can be absorbed by redefining the masses as follows
		\begin{equation}
			\tilde{m}_A=m_{A0}(1+\delta_A)
		\end{equation}
		Then, the Lagrangian of a particle test reads, at the Newtonian approximation
		\begin{align}
			L_T&=-\tilde{m}_T(1-\delta_T)c^2+\tilde{m}_T\frac{v_T^2}{2}(1-\delta_T)\nonumber\\
			&\quad+\sum_{A\ne T}\frac{\tilde{G}\tilde{m}_A\tilde{m}_T}{r_{AT}}(1+\delta_{AT})+O(c^{-2})+O(\delta_i^2) \label{eq_lagnewtdil}
		\end{align}
		The equations of motion read, at the Newtonian order
		\begin{equation}
			\bm{a}_T=\sum_{A\ne T}\tilde{G}\tilde{m}_A\frac{\bm{r}_{AT}}{r_{AT}^3}(1+\delta_T+\delta_{AT})
		\end{equation}
		At this step, we note that the weak equivalence principle is broken at several levels. The variation of the ratio between the inertial mass and the gravitational mass is encoded in the parameter $\delta_T$. But here a new violation appears through the parameter $\delta_{AT}$. In some cases, it is possible that $\alpha_0=0$ such that $\delta_T=0$ and only $\delta_{AT}\ne 0$, such that the weak equivalence principle violation cannot be reduced to a variation of the inertial/gravitational masses ratio \cite{minazzoli2016prd,viswanathan2018mn}.
		Let us note that only the scalars $\mu_A=\tilde{G}\tilde{m}_A$ can be measured by a gravitation experiment. We can multiply $L_T$ by $\tilde{G}$ and then only the scalars $\mu_A$ appear.
		At the post-Newtonian level, we can neglect the terms of order $O(\delta_T c^{-2})$ and $O(\delta_{AT}c^{-2})$. Indeed, if they exist, they will be detected first at the Newtonian level. Thus we keep only the universal coupling constant in the post-Newtonian approximation, since they have been absorbed at the Newtonian level by a unobservable redefinition of masses and gravitational constant. The resulting Lagrangian reads
		\begin{align}
			L_T&=-\mu_T(1-\delta_T)c^2+\frac{\mu_Tv_T^2}{2}(1-\delta_T)\nonumber\\
			&\quad+\sum_{A\ne T}\frac{\mu_A\mu_T}{r_{AT}}(1+\delta_{AT})+\frac{\mu_Tv_T^4}{8c^2}\nonumber\\
			&\quad+\sum_{A\ne T}\frac{\mu_A\mu_T}{r_{AT}c^2}\[ -2(\gamma+1)\bm{v}_T\cdot\bm{v}_A+\frac{2\gamma+1}{2}v_T^2\right.\nonumber\\
			&\left.\phantom{\sum_A}+(\gamma+1)v_A^2-\frac{\bm{r}_{AT}\cdot\bm{a}_A}{2}-\frac{(\bm{r}_{AT}\cdot\bm{v}_A)^2}{2r_{AT}^2} \] \nonumber\\
			&\quad-\sum_{A\ne T} \frac{\mu_A}{r_{AT}}\[\sum_{B\ne T}\frac{2\beta_T-1}{2c^2}\frac{\mu_B}{r_{BT}}+\sum_{B\ne A}\frac{2\beta_A-1}{c^2}\frac{\mu_B}{r_{AB}})\]\nonumber\\
			&\quad+ O(c^{-4})+O(c^{-2}\delta_i) + O(\delta_i^2) \label{eq_lagrangeT}
		\end{align}
		where
		\begin{equation}
			\gamma=\frac{1-\alpha_0^2}{1+\alpha_0^2}
		\end{equation}
		and where $\beta_A$ can be decomposed in a universal and non-universal part
		\begin{equation}
			\beta_A=1+\frac{\beta_0+\tilde{\beta}_A}{2}\frac{\alpha_0^2}{(1+\alpha_0^2)^2}=\beta+d\beta_A\label{eq_beta_A}
		\end{equation}
		where $\beta=1+\beta_0\alpha_0^2/2(1+\alpha_0^2)^2$ and $d\beta_A=\tilde{\beta_A}\alpha_0^2/2(1+\alpha_0^2)^2$.
		We can make the post-Newtonian expansion of the non-universal part of $\beta_A$ since it does not appear in the Newtonian part.
		We find the well known $\gamma$ and $\beta$ post-Newtonian parameters \cite{will2018book,klioner2000prd}, to which we add non-universal coupling constants $\delta_A$, $\delta_{AB}$, et $d\beta_A$. The derivation of this Lagrangian leads to the modified Einstein-Infeld-Hoffmann-Droste-Lorentz (EIHDL) equations, which are Eq. \eqref{eq_eihmod}.

\section{Global Lagrangian and Hamiltonian formulation, and first integrals}
\label{ap_lagr}
We give a global Lagrangian and Hamiltonian formulation of the modified EIHDL equations of motion in massless dilaton theory. Then we give an explicit form of the Euler-Lagrange equation in post-Newtonian formalism. 

	\subsection{Global Lagrangian formulation}
		We can write the global post-Newtonian Lagrangian of a $N$ body system. The total Lagrange function $L$ is not equal to the sum of the one-body Lagrangians. To get the same equations of motion, we need to check that $\partial L/\partial \bm{r}_A = \partial L_A/\partial \bm{r}|_{\bm{r}=\bm{r}_A}$. 
		To do so, one just has to summ $L_T$ over $T$ then symmetrize the explicit expressions including $A$ and $T$ interms of $\bm{z}_A$ and $\bm{v}_A$ \cite{damour1990prd}. 
		Moreover, we can simplify the Lagrangian and making disappear the term $\bm{a}_A$ by remarking that
		\begin{align}
			-\frac{\bm{r}_{AT}}{r_{AT}}\cdot\bm{a}_A&=-\frac{\d}{\d t}\(\frac{\bm{r}_{AT}}{r_{AT}}\cdot\bm{v}_A\)+\frac{(\bm{v}_T-\bm{v}_A)}{r_{AT}}\cdot\bm{v}_A\nonumber\\
			&\quad-\frac{(\bm{r}_{AT}\cdot\bm{v}_A)}{r_{AT}}\frac{(\bm{r}_{AT}\cdot\bm{v}_T)}{r_{AT}}+\(\frac{\bm{r}_{AT}\cdot\bm{v}_A}{r_{AT}}\)^2
		\end{align}
		One can then replace the term containing the acceleration by the right-hand-side ignoring the total derivative.
		We can also get the Lagrangian from the global Lagrangian of Damour \& Esposito-Far\`ese \cite{damour1992cqg} in the case of a conformal coupling, by substitutind our redefinitions of the constants. The result is the following -- we have replaced label $T$ by label $A$ in order to get a more ``canonical'' expression of the Lagrangian, according to ``canonical'' post-Newtonian litterature, for example \cite{will2018book,damour1990prd,klioner2000prd}-- :
		\begin{align}
			L&=-\sum_A\mu_A(1-\delta_A)c^2 + L_N + \frac{1}{c^2}\sum_A L_A \nonumber\\
			&\quad+ \frac{1}{2c^2}\sum_A\sum_{B\ne A}L_{AB} + \frac{1}{2c^2}\sum_A\sum_{B\ne A}\sum_{C\ne A}L^A_{BC} \label{eq_lagrange_dil}
		\end{align}
		where
		\begin{equation}
			L_N=\sum_A\mu_A(1-\delta_A)\frac{v_A^2}{2} + \frac{1}{2}\sum_A\frac{\mu_A\mu_B}{r_{AB}}(1+\delta_{AB})
		\end{equation}
		\begin{align}
			L_A=\frac{\mu_Av_A^4}{8},
		\end{align}
		\begin{align}
			L_{AB}&=\frac{\mu_A\mu_B}{r_{AB}}\left[ (2\gamma+1)v_A^2 -\frac{4\gamma+3}{2}\bm{v}_A\cdot\bm{v}_B\right.\nonumber\\
			&\quad\quad\quad\quad\quad\left.-\frac{1}{2}(\bm{v}_A\cdot\bm{n}_{AB})(\bm{v}_B\cdot\bm{n}_{AB}) \right]
		\end{align}
		and
		\begin{equation}
			L^A_{BC}=-\frac{\mu_A\mu_B\mu_C}{r_{AB}r_{AC}}(2\beta+2d\beta^A - 1) \label{eq_labc}
		\end{equation}
		where $\bm{n}_{AB}=\bm{r}_{AB}/r_{AB}$. 
		The first mass term $-\sum_A\mu_A(1-\delta_A)c^2$ 
		is useless to derive the equations of motion but will be usefull to express simply the first integrals, in particular the barycenter. The Newtonian part $L_N$ of the Lagrange function express the weak equivalence principle violation because, first, the inertial masses $\mu_A(1-\delta_A)$ are not equal to the gravitational masses, which become inseparable because their meaning is expressed only by the interaction of two bodies and are expressed by $\mu_A\mu_B(1+\delta_{AB})$. 
		The last three body term $L^A_{BC}$
		is also responsible for a weak equivalence principle violation because of $d\beta^A$, 
		which expresses that the three body interaction depends on the internal composition of the bodywhich is in motion but also of the two bodies which generate the gravitational in which the body is in motion. Indeed, when we derive this term with respect to $\bm{r}_T$, terms proportionnal to $d\beta^T$ appear but also terms proportional to $d\beta^A$ where $A\ne T$.  
		We note finally that $L_{AB}$ is symetric by permutation of $A$ and $B$, but $L^A_{BC}$ is only symetric by permutation of $B$ and $C$. This is because of these symetries that we had to divide the sums by two.

		Even by violating the equivalence principle by all these ways, this Lagrange function conserves the same symetries that the modified GR with post-Newtonian parameters $\beta$ and $\gamma$ \cite{klioner2016parametrized}. 
		In principle, if the massless dilaton theory was well respected, each mass should be modified but here we neglect the terms at the order $O(c^{-2}\delta_A)$.

		The linear momentum of body $A$ is
		\begin{align}
			\bm{p}_A &= \mu_A(1-\delta_A)\bm{v}_A + \frac{\mu_Av_A^2}{2c^2}\bm{v}_A \nonumber\\
			&\quad + \frac{1}{c^2}\sum_{B\ne A}\frac{\mu_A\mu_B}{r_{AB}}\left[ (2\gamma+1)\bm{v}_A-\frac{4\gamma+3}{2}\bm{v}_B\right.\nonumber\\
			&\quad\quad\quad\quad\quad\quad\quad\quad\left.-\frac{1}{2}(\bm{v}_B\cdot\bm{n}_{AB})\bm{n}_{AB} \right] \label{moment_A_dil}
		\end{align}
		The Lagrangian \eqref{eq_lagrange_dil} is invariant by spatial rotation and translation, and by temporal translation. The seven classical first integrals are well conserved. 
		Linear momentum:
		\begin{align}
			\bm{P}&=\sum_A\bm{p}_A \nonumber\\
			&=\sum_A\mu_A\bm{v}_A\left[ 1-\delta_A + \frac{1}{2c^2}\left( v_A^2 - \sum_{B\ne A}\frac{\mu_B}{r_{AB}} \right) \right] \nonumber\\
			&\quad- \frac{1}{2c^2}\sum_A\sum_{B\ne A}\frac{\mu_A\mu_B}{r_{AB}}(\bm{n}_{AB}\cdot\bm{v}_A)\bm{n}_{AB} \label{eq_momlin}
		\end{align}
		angular momentum:
		\begin{align}
			\bm{J}&=\sum_A\bm{z}_A\times\bm{p}_A\nonumber\\
			&=\sum_A\mu_A\bm{z}_A\times\bm{v}_A\left(1-\delta_A+\frac{v_A^2}{2c^2}\right) \nonumber\\
			&\quad +\frac{1}{c^2}\sum_A\sum_{B\ne A}\frac{\mu_A\mu_B}{r_{AB}}\left[ (2\gamma+1)\bm{z}_A\times\bm{v}_A \phantom{\frac{1}{1}}\right.\nonumber\\
			&\quad\left.- \frac{4\gamma+3}{2}\bm{z}_A\times\bm{v}_B-\frac{1}{2}(\bm{v}_B\cdot\bm{n}_{AB})\bm{z}_A\times\bm{n}_{AB} \right] \label{eq_momang}
		\end{align}
		energy:
		\begin{align}
			h&= \sum_A \bm{p}_A\cdot\bm{v}_A - L \nonumber\\
				 &=\sum_A(1-\delta_A)\mu_A\left(c^2+\frac{v_A^2}{2}\right)+ \sum_A\frac{3\mu_Av_A^4}{8c^2} \nonumber\\
				 &\quad+\frac{1}{2c^2}\sum_A\sum_{B\ne A}\sum_{C\ne A}\frac{\mu_A\mu_B\mu_C}{x_{AB}x_{AC}}(2\beta+2d\beta^A-1) \nonumber \\
				 &\phantom{=}-\frac{1}{2}\sum_A\sum_{B\ne A}\frac{\mu_A\mu_B}{x_{AB}}\left[ 1+\delta_{AB}-\frac{2\gamma+1}{c^2}v_A^2 \right.\nonumber\\
				 &\quad\left.+\frac{4\gamma+3}{2c^2}\bm{v}_A\cdot\bm{v}_B+\frac{1}{2c^2}(\bm{v}_A\cdot\bm{n}_{AB})(\bm{v}_B\cdot\bm{n}_{AB})\right] \label{energy_veloc}
		\end{align}
		\begin{sloppypar}
		Three more first integrals can be obtained -- they actually correspond to the Lorentz invariance. A direct derivation shows that the following vector is a first integral:
		\end{sloppypar}
		\begin{equation}
			\bm{q}=\bm{G}-\bm{V}t \label{const_lag_dil}
		\end{equation}
		where
		\begin{equation}
			\bm{G}=\frac{c^2}{h}\sum_A\mu_A\bm{z}_A\left(1-\delta_A+\frac{v_A^2}{2c^2}-\frac{1}{2c^2}\sum_{B\ne A}\frac{\mu_B}{r_{AB}}\right) \label{bary_lag_dil}
		\end{equation}
		are the coordinates of the relativistic barycenter of the system and
		\begin{equation}
			\bm{V}=\frac{c^2\bm{P}}{h}
		\end{equation}
		is the velocity of the barycenter motion. $\bm{q}$ is called barycenter constant, since we have
		\begin{equation}
			\bm{G}=\bm{q}+\bm{V}t
		\end{equation}
		and that $\bm{q}$ is the constant component of $\bm{G}$. 

	\subsection{Derivation of Euler-Lagrange equations of motion}	
		In order to avoid indexes confusion when we derive, we will derive $L$ with respect to $\bm{z}_X$ and $\bm{v}_X=\d\bm{z}_X/\d t$. This work use the same method that \cite{klioner2016parametrized} but is a generalization in the massless dilaton framework.

		Euler-Lagrange equations of motion write
		\begin{align}
			\bm{a}_X&=F_X^{-1}\left[ \bm{H}_X -\frac{1}{c^2}(\bm{v}_X\cdot\bm{a}_X)\bm{v}_X\right. \nonumber\\
			&\quad+\frac{1}{c^2}\sum_{B\ne X}\frac{\mu_B}{r_{XB}}\left( \frac{4\gamma+3}{2}\bm{a}_B\right. \nonumber\\
			&\quad\quad\quad\left.\left.+\frac{1}{2}(\bm{a}_B\cdot\bm{n}_{XB})\bm{n}_{XB}\right)\right] \label{ugly_lagrange}
		\end{align}
		where
		\begin{equation}
			F_X = 1-\delta_X+\frac{v_X^2}{2c^2}+\frac{2\gamma+1}{c^2}\sum_{B\ne X}\frac{\mu_B}{r_{XB}}
		\end{equation}
		and
		\begin{align}
			\bm{H}_X&=\sum_{B\ne X}\mu_B\frac{\bm{r}_{XB}}{r_{XB}^3}\left[1+\delta_{BX} - \frac{3}{2}(\bm{v}_B\cdot\bm{n}_{XB})^2\right. \nonumber\\
			&\quad\left.-(2\gamma+2)\bm{v}_X\cdot\bm{v}_B +(\gamma+1)v_B^2+\frac{2\gamma+1}{2}v_X^2\right] \nonumber\\
			&\quad-\frac{1}{c^2}\sum_{B\ne X}\frac{\mu_B}{r_{XB}^2}\left[ (2\gamma+1)(\bm{n}_{XB}\cdot(\bm{v}_X-\bm{v}_B))(\bm{v}_X-\bm{v}_B)\right. \nonumber\\
			&\phantom{espaceespaceespace}\left.-(\bm{n}_{XB}\cdot\bm{v}_B)\bm{v}_B \right]\nonumber\\
			&\quad-\frac{1}{c^2}\sum_{B\ne X}\mu_B\frac{\bm{r}_{XB}}{r_{XB}^3}\left( (2\beta+2d\beta^X-1)\sum_{C\ne X}\frac{\mu_C}{r_{XC}}\right. \nonumber\\
			&\quad\quad\quad\left.+(2\beta+2d\beta^B-1)\sum_{C\ne B}\frac{\mu_C}{r_{BC}} \right)
		\end{align}

		We note that like in \cite{klioner2016parametrized}, the exact resolution of Euler-Lagrange equations \eqref{ugly_lagrange} conserves exactly the following first integrals: linear momentum (Eq. \eqref{eq_momlin}), angular momentum (Eq. \eqref{eq_momang}), and energy (Eq. \eqref{energy_veloc}). 
		However, they are not easy to integrate because they do not appear in the form of an ordinary differential equation. To solve it numerically, one has to invert a matrix on each step of time. 
		However, in the first post-Newtonian approximation $O(c^{-2})$, one can get a second order ordinary differential equation but we loose the property of exactitude of the first integrals, a smooth signal remains at the order $O(c^{-4})$. By setting
		\begin{equation}
			F_X^{-1}=1+\delta_X-\frac{v_X^2}{2c^2}-\frac{2\gamma+1}{c^2}\sum_{B\ne X}\frac{\mu_B}{r_{XB}}+O(c^{-4})
		\end{equation}
		and by neglecting all the second post-Newstonian terms at order $O(c^{-4})$ that appear when the products are developped, one finds well EIHDL modified equations \eqref{eq_eihmod}.

	\subsection{Global Hamiltonian formulation}

		To get the global Hamiltonian, we perform a Legendre transformation. In this framework, we express energy  \eqref{energy_veloc} with respect to the conjugated variables $(\bm{z}_A,\bm{p}_A)$ instead of $(\bm{z}_A,\bm{v}_A)$. To do so, we have to invert Eq. \eqref{moment_A_dil} 
		which is always possible by perturbation at order $O(c^{-2})$, 
		then by substituting in Eq. \eqref{energy_veloc}, 
		still neglecting terms at order $O(c^{-4})$, $O(\delta_A^2)$ and $O(c^{-2}\delta_A)$. The result is:
		\begin{align} 
			\mathcal{H}&=\sum_A\left(\mu_A(1-\delta_A)c^2+\frac{p_A^2}{2\mu_A}(1+\delta_A)-\frac{p_A^4}{8\mu_A^3c^2}\right)\nonumber\\
			&\quad-\frac{1}{2}\sum_A\sum_{B\ne A}\frac{1}{x_{AB}}\left( \mu_A\mu_B(1+\delta_{AB})  \phantom{\frac{1}{1}} \right.\nonumber\\
			&\quad+\frac{1}{c^2}\frac{\mu_B}{\mu_A}(2\gamma+1)p_A^2  - \frac{4\gamma+3}{2c^2}\bm{p}_A\cdot\bm{p}_B \phantom{\frac{1}{1}}\nonumber\\
			&\quad\left. -\frac{1}{2c^2}(\bm{p}_A\cdot\bm{n}_{AB}) (\bm{p}_B\cdot\bm{n}_{AB})\right) \nonumber\\
			&\quad+\frac{1}{2c^2}\sum_A\sum_{B\ne A}\sum_{C\ne A}\frac{\mu_A\mu_B\mu_C}{x_{AB}x_{AC}}(2\beta+2d\beta^A-1) \label{hamiltonian}
		\end{align}
		The mass term $\sum_A \mu_A(1-\delta_A)c^2$ is useless for derivating equations of motion but will be usefull for the first integrals.
		The newtonian part of this Hamiltonian writes
		\begin{equation}
			\mathcal{H}_N = \sum_A\frac{p_A^2}{2\mu_A}(1+\delta_A)-\frac{1}{2}\sum_A\sum_{B\ne A}\frac{\mu_A\mu_B}{r_{AB}}(1+\delta_{AB}).
		\end{equation}
		In the conjugated variables, the first integrals have simpler expressions.
		Linear momentum:
		\begin{equation}
			\bm{P}=\sum_A\bm{p}_A
		\end{equation}
		angular momentum:
		\begin{equation}
			\bm{J}=\sum_A\bm{z}_A\times\bm{p}_A
		\end{equation}
		the energy is $\mathcal{H}$ itself, and the barycenter constant:
		\begin{equation}
			\bm{q}=\bm{G}-\bm{V}t \label{const_ham_dil}
		\end{equation}
		where
		\begin{equation}
			\bm{G}=\frac{c^2}{\mathcal{H}}\sum_A\mu_A\bm{z}_A\left(1-\delta_A+\frac{p_A^2}{2\mu_Ac^2}-\frac{1}{2c^2}\sum_{B\ne A}\frac{\mu_B}{r_{AB}}\right) \label{bary_ham_dil}
		\end{equation}
		and
		\begin{equation}
			\bm{V}=\frac{c^2\bm{P}}{\mathcal{H}}
		\end{equation}
        Here, as in the Lagrangian formalism, linear momentum, angular momentum and energy are exactly conserved when Hamilton equations of motion 
		\begin{equation}
			\frac{\d \bm{z}_A}{\d t}=\frac{\partial \mathcal{H}}{\partial \bm{p}_A},\quad 
			\frac{\d \bm{p}_A}{\d t}=-\frac{\partial \mathcal{H}}{\partial \bm{z}_A}
		\end{equation}
        are integrated exactly.

\section{Nordtvedt effect in light dilaton framework}
\label{app_nor}
    We derive the Nordtvedt effect in the dilaton framework in order to derive eq. \eqref{eq_noreffect}.
	As did Damour \& Esposito-Far\`ese \cite{damour1992cqg}, the idea consists in introducing a sensitivity parameter
	\begin{equation}
		s_A=-\frac{\partial \ln m_A}{\partial\ln G_L}
	\end{equation}
	where $G_L$ is the locally measured gravitational constant. In the weak-field limit, this sensitivity is $s_A=|\Omega_A|/m_Ac^2$ where
	\begin{equation}
		\Omega_A=G\int_A\int_A\frac{\rho_A(\bm{r})\rho_A(\bm{r}')}{|\bm{r}-\bm{r}'|}\ \d^3r\d^3r'
	\end{equation}
	is the auto gravitation energy. In our parametrization, $\tilde{\alpha}_A$ should be modified as
	\begin{equation}
		\tilde{\alpha}_A'=\frac{\partial\ln \tilde{m}_A}{\partial\phi}+\frac{\partial \ln{m}_A}{\partial\ln\tilde{G}}\frac{\d\ln\tilde{G}}{\d\phi}=\tilde{\alpha}_A-\frac{|\Omega_A|}{\tilde{m}_Ac^2}\frac{\d\ln\tilde{G}}{\d\phi}
	\end{equation}
	where $\tilde{\alpha}_A$ is the coefficient already computed with respect to the dilatonic charges. We have also
	\begin{equation}
		\tilde{G}=\frac{G}{f(\phi_0)}(1+\alpha_u^*(\phi_0)^2)
	\end{equation}
	thus
	\begin{equation}
		\frac{\d\ln\tilde{G}}{\d\phi}=-\sqrt{\frac{2}{Z(\varphi)}}\frac{f'(\varphi)}{f(\varphi)}+\frac{2\alpha_0\beta_0}{1+\alpha_0^2}
	\end{equation}
	To find back the classical parametrization, we can redefine the constant without measurable modification 
	\begin{equation}
		\tilde{G}\mapsto \e^{2D(\varphi)}\tilde{G}
	\end{equation}
	\begin{equation}
		\tilde{m}_A\mapsto \e^{D(\varphi)}\tilde{m}_A
	\end{equation}
	where
	\begin{equation}
		D(\varphi)=\int_{\varphi_0}^\varphi\alpha_u(x)\d x
	\end{equation}
	Then
	\begin{align}
		\left.\frac{\d\ln\tilde{G}}{\d\phi}\right|_{\phi_0}&=\sqrt{\frac{2}{Z(\varphi_0)}}\(2\alpha_u(\varphi_0)-\frac{f'(\varphi_0)}{f_0}\)+\frac{2\alpha_0\beta_0}{1+\alpha_0^2}\nonumber\\
		&=2\alpha_0+\frac{2\alpha_0\beta_0}{1+\alpha_0^2}
	\end{align}
	On the other hand, we have
	\begin{equation}
		\frac{\partial \ln\tilde{m}_A}{\partial \phi}=\tilde{\alpha}_A+\frac{\d\delta_A/\d\phi}{1+\delta_A}
	\end{equation}
	The absence of $\alpha_0$ comes from the redefinition $\tilde{m}_A\mapsto \e^{-D(\phi)}\tilde{m}_A$.  
	We have then
	\begin{align}
		\delta_A'&=\frac{\alpha_0\tilde{\alpha}_A'}{1+\alpha_0^2}\\
		&=\frac{\alpha_0\tilde{\alpha}_A}{1+\alpha_0^2}-\[2\frac{\alpha_0^2}{1+\alpha_0^2}+\frac{2\alpha_0^2\beta_0}{(1+\alpha_0^2)^2}\]\frac{|\Omega_A|}{\tilde{m}_Ac^2}\\
		&=\delta_A-(4\beta-\gamma-3)\frac{|\Omega_A|}{\tilde{m}_Ac^2}
	\end{align}
	where $\delta_A$ is the deviation from the weak equivalence principle. We have neglected the term $\alpha_0\d\delta_A/\d\phi/((1+\delta_A)(1+\alpha_0^2))$ because it would make appear terms of order three in $\alpha_i$. For roughly spherical bodies, we can estimate $\Omega_A$ 
	\begin{align}
		\frac{\Omega_A}{\tilde{m}_Ac^2}&=-\frac{\tilde{G}}{2\tilde{m}_Ac^2}\int_A\int_A \frac{\rho(\bm{x})\rho(\bm{x}')}{\|\bm{x}-\bm{x}'\|}\d^3x\d^3x'+O(c^{-2})\\
		&\approx-\frac{3\mu_A}{5R_Ac^2}
	\end{align}
	where $R_A$ is the average radius of the body. We can limit our calculation to this approximation until we have a positive detection of the strong equivalence principle violation.

\end{document}